\documentclass[fleqn,11pt]{article}
\usepackage{amsfonts,epsfig}
\usepackage{times}

\newcommand{\Ga}{\alpha}
\newcommand{\Gb}{\beta}

\newcommand{\Gd}{\delta}
\newcommand{\Ge}{\epsilon}
\newcommand{\Gg}{\gamma}
\newcommand{\GG}{\Gamma}
\newcommand{\Gk}{\kappa}
\newcommand{\Gl}{\lambda}
\newcommand{\GL}{\Lambda}

\newcommand{\Go}{\omega}

\newcommand{\GS}{\Sigma}

\newcommand{\Gth}{\theta}
\newcommand{\GTh}{\Theta}

\newcommand{\cA}{{\scriptscriptstyle\cal A}}
\newcommand{\cB}{{\scriptscriptstyle\cal B}}
\newcommand{\cC}{{\scriptscriptstyle\cal C}}
\newcommand{\cD}{{\scriptscriptstyle\cal D}}

\newcommand{\cM}{{\scriptscriptstyle\cal M}}
\newcommand{\cN}{{\scriptscriptstyle\cal N}}
\newcommand{\cK}{{\scriptscriptstyle\cal K}}
\newcommand{\cL}{{\scriptscriptstyle\cal L}}
\newcommand{\cP}{{\scriptscriptstyle\cal P}}

\newcommand{\CA}{{\cal A}}
\newcommand{\CB}{{\cal B}}

\newcommand{\CD}{{\cal D}}

\newcommand{\CM}{{\cal M}}
\newcommand{\CN}{{\cal N}}

\newcommand{\CL}{{\cal L}}
\newcommand{\CP}{{\cal P}}
\newcommand{\CQ}{{\cal Q}}

\newcommand{\cJ}{{\cal J}}
\newcommand{\cO}{{\cal O}}

\newcommand{\cV}{{\cal V}}

\newcommand{\dA}{{\dot{A}}}
\newcommand{\dB}{{\dot{B}}}
\newcommand{\dC}{{\dot{C}}}
\newcommand{\dD}{{\dot{D}}}

\newcommand{\Beps}{\overline{\Ge}}
\newcommand{\Bchi}{\overline{\chi}}
\newcommand{\Bpsi}{\overline{\psi}}

\newcommand{\TA}{{\tilde A}}

\newcommand{\ft}[2]{{\textstyle {\frac{#1}{#2}} }}

\newcommand{\dd}{\partial}

\newcommand{\ra}{\rightarrow}
\newcommand{\I}{{\rm i}}

\newcommand{\be}{\begin{equation}}
\newcommand{\ee}{\end{equation}}
\newcommand{\ben}{\begin{displaymath}}
\newcommand{\een}{\end{displaymath}}
\newcommand{\ba}{\begin{eqnarray}}
\newcommand{\ea}{\end{eqnarray}}
\newcommand{\nn}{\nonumber}
\newcommand{\non}{\nonumber\\}

\newcommand{\mathon}{\mathversion{bold}}
\newcommand{\mathoff}{\mathversion{normal}}

\makeatletter
\@addtoreset{equation}{section}
\makeatother
\renewcommand{\theequation}{\thesection.\arabic{equation}}

\newcommand{\la}{\label}
\newcommand{\ci}{\cite}

\newcommand{\Ref}[1]{(\ref{#1})}

\def\moth{\mathsurround=0pt}
\newdimen\zo \zo=0pt

\def\tick{\leaders\hrule height 0.5ex depth 0pt \hskip 0.5pt}
\def\upboxfill{$\moth \setbox\zo\hbox{\tick}%
  \hskip 2pt\hbox to 0pt{$\tick$\hss}\hrulefill \hbox to 6pt{$\tick$\hss}$}
\def\underbox#1{\offinterlineskip{\mathord{\mathop{\vtop{\moth\ialign{##\crcr
      $\hfil\displaystyle{#1}\hfil$\crcr\noalign{}
      {\upboxfill}\crcr\noalign{}}}}\limits}}}
\def\dtick{\leaders\hrule height .34pt depth .5ex \hskip 0.5pt}
\def\downboxfill{$\moth \setbox\zo\hbox{\dtick}%
  \hskip 2pt\hbox to 0pt{$\dtick$\hss}\hrulefill \hbox to 6pt{$\dtick$\hss}$}
\def\overbox#1{\mathop{\vbox{\moth\ialign{##\crcr\noalign{}
\downboxfill\crcr\noalign{\vskip 1pt\nointerlineskip}
      $\hfil\displaystyle{#1}\hfil$\crcr}}}\limits}

\def\undersym#1{\underbox{{}#1}}
\def\oversym#1{\overbox{{}#1}}

\newcommand{\vl}{{\vphantom{[}}}
\newcommand{\VV}[2]{{\cV^{\,#1}{}\!^\vl_{#2}}}
\newcommand{\TT}[2]{{T^\vl_{#1|#2}}}
\newcommand{\TTd}[3]{{T^{\,\bf #1}_{#2|#3}}}

\newcommand{\EE}{E_{8(8)}}

\newcommand{\mfg}{{\mathfrak g}}

\newcommand{\ovi}{{\overline{i}}}
\newcommand{\ovj}{{\overline{j}}}
\newcommand{\ovk}{{\overline{k}}}
\newcommand{\ovl}{{\overline{l}}}

\newcommand{\cro}{\!\times\!}
\newcommand{\equ}{\!=\!}

\newcommand{\cc}[1]{g_{\scriptscriptstyle #1}}
\newcommand{\dimg}{\nu}

\begin{document}

\thispagestyle{empty}

\begin{flushright}
AEI-2001-017\\
LPTENS-01/13
\end{flushright}
\renewcommand{\thefootnote}{\fnsymbol{footnote}}

\vspace*{0.1ex}
\begin{center}
{\bf\LARGE Compact and Noncompact Gauged}
\bigskip

{\bf\LARGE Maximal Supergravities in Three Dimensions}

\bigskip\bigskip\medskip

{\bf H.~Nicolai\medskip\\ }
{\em Max-Planck-Institut f{\"u}r Gravitationsphysik,\\
  Albert-Einstein-Institut,
\footnote{Supported in part by the European Union under Contracts
No. HPRN-CT-2000-00122 and No. HPRN-CT-2000-00131.}\\
  M\"uhlenberg 1, D-14476 Potsdam, Germany}

\smallskip {\small nicolai@aei-potsdam.mpg.de} \bigskip\smallskip
\addtocounter{footnote}{-1}

{\bf H.~Samtleben\medskip\\ }
{\em Laboratoire de Physique Th{\'e}orique\\
  de l'{\'E}cole Normale Sup{\'e}rieure,
\footnotemark $^,$\footnote{UMR 8549: Unit{\'e} Mixte du Centre
National de la Recherche Scientifique, et de l'{\'E}cole Normale
Sup{\'e}rieure. }\\ 
24 Rue Lhomond, F-75231 Paris Cedex 05, France}~
\\

\smallskip {\small henning@lpt.ens.fr\medskip} 
\end{center}
\renewcommand{\thefootnote}{\arabic{footnote}}
\setcounter{footnote}{0}
\bigskip

\begin{abstract}
We present the maximally supersymmetric three-dimensional gauged
supergravities. Owing to the special properties of three dimensions
--- especially the on-shell duality between vector and scalar fields,
and the purely topological character of (super)gravity --- they
exhibit an even richer structure than the gauged supergravities in
higher dimensions. The allowed gauge groups are subgroups of the
global $E_{8(8)}$ symmetry of ungauged $N\equ16$ supergravity. They
include the regular series $SO(p,8\!-\!p)\!\times\!SO(p,8\!-\!p)$ for
all $p\equ0, 1, \dots, 4$, the group $E_{8(8)}$ itself, as well as
various noncompact forms of the exceptional groups $E_7, E_6$ and
$F_4\cro G_2$. We show that all these theories admit maximally
supersymmetric ground states, and determine their background
isometries, which are superextensions of the anti-de Sitter group
$SO(2,2)$. The very existence of these theories is argued to point to
a new supergravity beyond the standard $D\!=\!11$ supergravity.
\end{abstract}

\renewcommand{\thefootnote}{\arabic{footnote}}
\vfill

\bigskip

\leftline{hep-th/0103032}

\leftline{{March 2001}}

\setcounter{footnote}{0}
\newpage

\section{Introduction}

In this article we explain in detail the construction of maximal
gauged supergravities in three dimensions, recently announced in
\cite{NicSam01}. While maximal gauged supergravities in higher
dimensions have been known for a long time, starting with the gauged
$N\!=\!8$ theory in four dimensions \ci{deWNic82}, and subsequently
for dimensions $5\!\leq D\!\leq \!8$
\ci{PePivN84,PePivN85,SalSez85,GuRoWa86,Cowd99}, the results on gauged
supergravities in three dimensions and below have remained somewhat
fragmentary until now. The results presented in this paper close this
gap.  In addition they open up new perspectives: unlike maximal gauged
supergravities in higher dimensions, the maximal AdS$_3$
supergravities, which we obtain here, are neither contained in nor
derivable by any known mechanism from the known maximal supergravities
in higher dimensions. The new, and purely field theoretic, evidence
for a theory beyond $D\!=\!11$ supergravity \cite{CrJuSc78} and type
IIB supergravity \cite{GreSch83,SchWes83} that we have thus obtained
is perhaps the most important consequence of the present work.

Topological gauged supergravities in three dimensions were first
constructed in \ci{AchTow86}; these theories are supersymmetric
extensions of Chern-Simons (CS) theories with $(n_L,n_R)$
supersymmetry and gauge group $SO(n_L)\!\times\!SO(n_R)$, but have no
propagating matter degrees of freedom (see also \ci{DesKay83} for
earlier work on $D\!=\!3$ supergravity). Matter coupled gauged
supergravities can, of course, be obtained by direct dimensional
reduction of gauged supergravities in $D\!\geq\!4$ to three dimensions
and below, but these do not preserve the maximal supersymmetry
\cite{LuPoTo97}.  Another matter coupled theory with half maximal
supersymmetry, obtained by compactifying the ten-dimensional $N\equ1$
supergravity on a seven-sphere, has been discussed in \cite{CvLuPo00}
(however, \cite{CvLuPo00} deals only with the bosonic part of the
Lagrangian). In a different vein, \cite{DKSS00} constructs an abelian
gauged supergravity by deforming the $D\equ3$, $N\equ2$ supergravity
whose matter sector is described by an $SO(n,2)/SO(n)\cro SO(2)$ coset
space sigma model. This model bears some resemblance to the present
work in that the vector fields appear via a CS term rather than a
Yang-Mills term, unlike the matter-coupled theories mentioned
before. However, the construction is limited to the abelian case,
whereas the present construction yields non-abelian CS theories,
thereby providing the first examples of a non-abelian duality between
scalars and vector fields in three space-time dimensions.

Gauged supergravities have attracted strong interest again recently in
the context of the conjectured duality between AdS supergravities and
superconformal quantum field theories on the AdS boundary
\cite{AGMOO00}.  For instance, classical supergravity domain wall
solutions are claimed to encode the information on the renormalization
group flow of the strongly coupled gauge theory \cite{FGPW99}. The
theories admitting AdS$_3$ ground states are expected to be of
particular interest for the AdS/CFT duality due to the rich and rather
well understood structure of two-dimensional superconformal field
theories. However, a large part of the recent work dealing with the
conjectured AdS/CFT correspondence in AdS$_3$ has been based on the
BTZ black hole solution of \cite{BaTeZa92}, which has no propagating
matter degrees of freedom in the bulk. We will see that the gauged 
$N\!=\!16$ theories yield a rich variety of supersymmetric groundstates,
virtually exhausting all the possible vacuum symmetries of AdS type
listed in \cite{GuSiTo86}, and thus an equally rich variety of
superconformal theories on the boundary.

As is well known \cite{CreJul79}, the scalar fields in the toroidal
compactification of $D\!=\!11$ supergravity \cite{CrJuSc78} on a
$d$-torus form a coset space sigma model manifold $G/H$ with the
exceptional group $G\equ E_{d(d)}$ and $H$ its maximally compact
subgroup; in particular, for $d\equ8$ one obtains a theory with global
$E_{8(8)}$ symmetry and local $SO(16)$ \cite{Juli83,MarSch83}. The 
complete list of ungauged matter coupled supergravities in three
dimensions (which unlike topological supergravities only exist with 
$N\leq 16$ supersymmetries) has been presented in \cite{dWToNi93}. Gauging 
any of these theories corresponds to promoting a subgroup $G_0$ of 
the rigid $G$ symmetry group to a local symmetry in such a way that
the full local supersymmetry is preserved. The latter requirement
engenders additional Yukawa-like couplings between the scalars and
fermions, as well as a rather complicated potential for the scalar
fields. As we will demonstrate by explicit construction, the possible
compact and non-compact non-abelian gauge groups, all of which are
subgroups of the global $\EE$ symmetry of the ungauged maximal
supergravity theory and preserve the full local $N\!=\!16$
supersymmetry, are more numerous in three dimensions than in higher
dimensions.

There are essentially two properties which distinguish the three
dimensional models from all their higher dimensional relatives. First,
the gravitational sector does not contain any propagating degrees of
freedom such that the theories without matter coupling may be
formulated as CS theories of AdS supergroups \ci{AchTow86}; see 
also the classic article \cite{DeJaHo84} for a description of the 
peculiarities of gravity in three space-time dimensions. In fact,
pure quantum gravity \cite{Witt88,AHRSS89} and quantum supergravity
\cite{dWMaNi93} are exactly solvable in three space-time dimensions.
Second, in three dimensions scalar fields are on-shell equivalent to 
vector fields. At the linearized level, this duality is encapsulated
in the relation
\be\la{dualityI}
\Ge_{\mu\nu\rho}\, \dd^\rho \varphi^m
=  \dd_{[\mu} {B_{\nu]}}^m \;.
\ee
This relation plays a special role in the derivation of maximal $N\!=\!16$ 
supergravity in three dimensions \ci{Juli83,MarSch83,CJLP98,Mizo98}: in
order to expose its rigid $\EE$ symmetry, all vector fields obtained
by dimensional reduction of $D\equ 11$ supergravity \ci{CrJuSc78} on
an 8-torus must be dualized into scalar fields. Vice versa, the duality 
\Ref{dualityI} allows us to redualize part of the scalar fields into 
vector fields, such that the ungauged theory possesses different 
equivalent formulations which are related by duality \cite{CJLP98}. 
As explained there, the replacement of scalar fields by vector fields 
breaks the exceptional $\EE$ symmetry; when attempting to gauge this 
theory while maintaining its $E_{8(8)}$ structure and thus keeping all 
the scalars, it is therefore {\it a priori} not clear how to re-incorporate
the vector fields necessary for the gauging without introducing new and 
unwanted propagating degrees of freedom. We will circumvent this apparent
problem by interpreting \Ref{dualityI} as defining up to 248 vector 
fields as (nonlocal) functions of the scalar fields. This freedom in 
the choice of the number of vector fields is at the origin of the large 
number of possible gauge groups that we encounter in three dimensions.

In higher dimensions, the gauge group is to a large extent determined
by the number and transformation behavior of the vector fields under
the rigid $G$ symmetry of the ungauged theory. As a necessary
condition for gauging a subgroup $G_0\subset G$, the vector fields or
at least a maximal subset thereof must transform in the adjoint
representation of $G_0$. In the latter case there may remain
additional vector fields which transform nontrivially under the gauge
group. Upon gauging, these charged vector fields would acquire mass
terms and thereby spoil the matching of bosonic and fermionic degrees
of freedom; to avoid such inconsistencies one needs some additional
mechanism to accommodate these degrees of freedom. Altogether, this
does not leave much freedom for the choice of the gauge group.  In
$D\equ 4$ and $D\equ 7$ one must make use of the full set of vector
fields transforming in the adjoint representation of the gauge groups
$SO(8)$ and $SO(5)$, respectively. The situation is more subtle in
dimensions $D\equ 5,6$ where only a subset of the vector fields
transforms in the adjoint representation of the gauge groups $SO(6)$
and $SO(5)$, respectively. The problem of coupling charged vector
fields is circumvented in $D\equ 5$ by dualizing the additional vector
fields into massive self-dual two forms \cite{PePivN85,GuRoWa86}; in
$D\equ6$ they are absorbed by massive gauge transformations of the
two forms \cite{Cowd99}.

By contrast the proper choice of gauge group is much less obvious in
three dimensions. With \Ref{dualityI}, we may introduce for any
subgroup $G_0\subset\EE$ a set of $\dimg={\rm dim~} G_0$ vector fields
transforming in the adjoint representation of $G_0$. A priori, there
is no restriction on the choice of $G_0$; however, demanding 
maximal supersymmetry of the gauged theory strongly restricts 
the possible choices for $G_0$. It is one of our main results that
the entire set of consistency conditions for the three-dimensional 
gauged theory may be encoded into a single algebraic condition
\be\la{critI}
{\mathbb P}_{\bf 27000}\;\GTh ~=~0 \;, 
\ee
where $\GTh$ is the embedding tensor characterizing the subgroup
$G_0$, and ${\mathbb P}$ a projector in the $\EE$ tensor product
decomposition $({\bf 248}\times{\bf 248})_{\rm sym} = {\bf 1}\!+\!{\bf
3875}\!+\!{\bf 27000}$. Solutions  to \Ref{critI} may be 
found by purely group theoretical
considerations. Having formulated the consistency conditions of the
gauged theory as a projector condition for the embedding tensor of the
gauge group allows us to construct a variety of models with
maximal local supersymmetry. As a result, we identify a ``regular''
series of gauged theories with gauge group $SO(p,8\!-\!p)\cro
SO(p,8\!-\!p)$, including the maximal compact gauge group $SO(8)\cro
SO(8)$ as a special case.  In addition, we find several theories with
exceptional noncompact gauge groups, among them an extremal theory
which gauges the full $\EE$ symmetry. These theories have no analog in
higher dimensions.

This collection of maximal admissible gauge groups is presented in
Table~\ref{ncgroups}; all the gauge groups --- apart from the
theory with local $\EE$ --- have two simple factors with a
fixed ratio of coupling constants. As a by-product of our construction
we can understand and re-state the corresponding consistency
conditions for the higher dimensional gauged supergravities of
\cite{deWNic82,GuRoWa86} in very simple terms; in particular, the
derivation of the $T$-identities for the $D=4,5$ theories can now be
simplified considerably by reducing it to purely group theoretical
condition analogous to \Ref{critI}. Remarkably, and even though the
rigid $G=E_{d(d)}$ symmetry of the ungauged theory is broken, the
construction and proof of consistency of the gauged theory makes
essential use of the properties of the maximal symmetry group
$E_{d(d)}$ in all cases.

\begin{table}[htb]
\centering
\vspace{2ex}
\begin{tabular}{||c|c||} 
\hline
&\\[-1.8ex]
gauge group $G_0$
&
ratio of coupling constants \\[.5ex]
\hline
\hline
&\\[-1.8ex]
$SO(p,8\!-\!p)\cro SO(p,8\!-\!p)$ 
& $g_1/g_2=-1$
\\[.5ex]
\hline
\hline
&\\[-1.8ex]
$G_{2(2)}\!\times\!F_{4(4)}$ 
&\\ [-1ex]
& $\cc{G_2}/\cc{F_4}=-3/2$ \\[-1ex] \cline{1-1}
&\\[-1.8ex]
$G_{2}\times F_{4(-20)}$ 
&\\[.5ex]
\hline
&\\[-1.8ex]
$E_{6(6)}\!\times\!SL(3)$
&\\[.5ex] \cline{1-1}
&\\[-1.8ex]
$E_{6(2)}\!\times\!SU(2,1)$
&
$\cc{A_2}/\cc{E_6}= -2$
\\[.5ex] \cline{1-1}
&\\[-1.8ex]
$E_{6(-14)}\!\times\!SU(3)$
&\\[.5ex] 
\hline
&\\[-1.8ex]
$E_{7(7)}\!\times\!SL(2)$ 
&\\ [-1ex]
& $\cc{A_1}/\cc{E_7}= -3 $\\[-1ex] \cline{1-1}
&\\[-1.8ex]
$E_{7(-5)}\!\times\!SU(2) $ 
&\\[.5ex]
\hline
&\\[-1.8ex]
$E_{8(8)}$
&
$\cc{E_8}$ \\[.5ex]
\hline
\end{tabular}
\bf\small
\caption{{\rm\small Regular and exceptional admissible gauge groups.}} 
\label{ncgroups}      
\end{table}

This paper is organized as follows. In Chapter~\ref{CHungauge} we
review the ungauged $N\equ 16$ theory and in particular discuss the
full nonlinear version of the duality \Ref{dualityI} between scalar
and vector fields. In Chapter~\ref{CHgauge} we present the Lagrangian
of the gauged theory. It is characterized by a set of tensors
$A_{1,2,3}$ which are nonlinear functions of the scalar fields and
describe the Yukawa-type couplings between fermions and scalars as
well as the scalar potential.  We derive the consistency conditions
that these tensors must satisfy in order for the full $N\!=\!16$
supersymmetry to be preserved, and show that $A_{1,2,3}$ combine into
a ``$T$-tensor'' analogous to the one introduced in \cite{deWNic82},
but now transforming as the ${\bf 1} + \bf{3875}$ of $E_{8(8)}$. In
Chapter~\ref{CHT} we show that these consistency conditions imply and
may entirely be encoded into the algebraic equation \Ref{critI} for
the embedding tensor of the gauge group, which selects the admissible
gauge groups $G_0\subset\EE$. In turn, every solution to \Ref{critI}
yields a nontrivial solution for $A_{1,2,3}$ in terms of the scalar
fields which satisfies the full set of consistency conditions. Maximal
supersymmetry of the gauged theory thus translates into a simple
projector equation for the gauge group $G_0$. 

In Chapter~\ref{CHG0} we analyze equation \Ref{critI} and its
solutions among the maximal subgroups of $SO(16)$ and $\EE$,
respectively. We find the maximal compact admissible gauge group
$G_0=SO(8)\cro SO(8)$ as well as its noncompact real forms
$SO(p,8\!-\!p)\cro SO(p,8\!-\!p)$ for $p\equ1,...,4$. In addition, we
identify the exceptional noncompact gauge groups given in
Table~\ref{ncgroups}. Each of these groups gives rise to a maximally
supersymmetric gauged supergravity.  Chapter~\ref{CHgs} is devoted to
an analysis of stationary points of the scalar potential which
preserve the maximal number of $16$ supersymmetries. We show that all
our theories admit a maximally symmetric ground state and determine
their background isometries. Finally we speculate on a possible higher
dimensional origin of these theories.

\mathon
\section{The ungauged $N\equ 16$ theory}
\mathoff
\la{CHungauge}

We first summarize the pertinent results about (ungauged) maximal
$N\equ 16$ supergravity in three dimensions. The complete Lagrangian
and supersymmetry transformations were presented in \ci{MarSch83},
whose conventions and notation we follow throughout this
paper.\footnote{In particular we use the metric with signature $(+--)$
and three-dimensional gamma matrices with $e\,\Gg^{\mu\nu\rho} =
- \I\Ge^{\mu\nu\rho}$, where $\Ge^{012}=\Ge_{012}=1$, and $e\equiv 
{\rm det} \, e_\mu{}^\alpha$ is the dreibein determinant.}  The physical
fields of $N\equ 16$ supergravity constitute an irreducible
supermultiplet with 128 bosons and 128 fermions transforming as
inequivalent fundamental spinors of $SO(16)$. In addition, the theory
contains the dreibein ${e_\mu}^\alpha$ and 16 gravitino fields
$\psi_\mu^I$, which do not carry propagating degrees of freedom in
three dimensions. As first shown in \cite{Juli83}, it possesses a
``hidden'' invariance under rigid $E_{8(8)}$ and local $SO(16)$
transformations. Consequently, the scalar fields are described by an
element $\cV$ of the non-compact coset space $E_{8(8)}/SO(16)$ in the
fundamental 248-dimensional representation of $E_{8(8)}$, which
transforms as
\be
\cV (x)\; \longrightarrow \; g\, \cV(x) \,h^{-1}(x)\;, \qquad
  g \in E_{8(8)} \; , \; h(x) \in SO(16) \;,
\label{GHV}
\ee
(see App.~A for our $\EE$ conventions). The scalar fields
couple to the fermions via the currents 
\be\la{VdV}
\cV^{-1} \dd_\mu \cV = \ft12 Q_\mu^{IJ} X^{IJ}\!+\!P_\mu^A Y^A \;.
\ee
The composite $SO(16)$ connection $Q_\mu^{IJ}$ enters the
covariant derivative $D_\mu$ in 
\ba
D^{\vphantom b}_\mu \psi_\nu^I &:=& 
\dd^{\vphantom b}_\mu \psi_\nu^I + 
\ft14\,\Go_\mu{}^{ab}\,\Gg_{ab}\, \psi_\nu^I  +
Q_\mu^{IJ} \psi_\nu^J 
\;,\non
D_\mu \chi^\dA &:=& \dd_\mu \chi^\dA + 
\ft14\,\Go_\mu{}^{ab}\,\Gg_{ab}\, \chi^\dA  +
\ft14\, Q_\mu^{IJ} \GG^{IJ}_{\dA\dB} \,\chi^\dB \;.
\ea
Definition \Ref{VdV} implies the integrability relations:
\be
Q_{\mu\nu}^{IJ}+\ft12\,\GG^{IJ}_{AB}\, P^A_\mu P^B_\nu = 0 \;,
\qquad
D^{\vphantom b}_{[\mu} P_{\nu]}^A = 0 \;,
\la{int}
\ee
where the $SO(16)$ field strength is defined as
\ben
Q_{\mu\nu}^{IJ}:= \dd_\mu Q_\nu^{IJ} - \dd_\nu Q_\mu^{IJ}
                  + 2 \,Q_\mu^{K[I} Q_\nu^{J]K}
\;.
\een
The full supersymmetry variations read \cite{MarSch83}
\ba\la{susyf}
\delta {e_\mu}^\Ga &=& 
\I \Beps^I \gamma^\Ga \psi_\mu^I 
\;,\qquad\quad
\delta\,\psi^I_\mu ~=~ D_{\mu}\epsilon^I - 
   \ft14 \I \Gg^\nu \Ge^J \, \Bchi \,\GG^{IJ} \Gg_{\mu\nu} \chi  
\;, 
\non[1ex]
\cV^{-1} \delta \cV &=& 
\Gamma^I_{A\dot A} \,\Bchi^{\dot A} \Ge^I Y^A
\;,\quad
\Gd\,\chi^\dA ~=~
\ft{\I}2\,\Gg^\mu\Ge^I\,\GG^I_{A\dA}\,\widehat{P}_\mu^A \;,
\ea
with the supercovariant current
\ben
\widehat{P}_\mu^A := P_\mu^A - \Bpsi_\mu^I \chi^\dA \GG^I_{A\dA} \;.
\een
As shown in \cite{MarSch83}, they leave invariant the
Lagrangian~\footnote{Note that the factor in front of the last 
term $(\Bchi\Gg_\mu\GG^{IJ}\chi)^2$ differs from the one given
in \cite{MarSch83} as was already noticed in \cite{Nico87a}.}
\ba\la{Lag}
\CL &=& 
 -\ft14 e R
 + \ft14 e P^{\mu A} P^A_\mu
+\ft12\, \Ge^{\Gl\mu\nu} {\Bpsi}{}^I_\Gl D_\mu \psi_\nu^I 
\non[1ex]
&&{}
-\ft{\I}{2}e {\Bchi}^{\dA} 
\Gg^\mu D_\mu \chi^{\dA} 
-\ft12e\,  {\Bchi}^{\dA} 
\Gg^\mu \Gg^\nu \psi^I_\mu \,\GG^I_{A\dA} P^A_\nu 
\non[1ex]
&& -\ft18 e\Big( \Bchi \Gg_\rho \GG^{IJ} 
\chi \left( \Bpsi{}_\mu^I
 \Gg^{\mu\nu\rho} \psi_\nu^J 
- \Bpsi{}_\mu^I \Gg^\rho \psi^{\mu J}\right)
+ \Bchi \chi \, \Bpsi{}_\mu^I \Gg^\nu \Gg^\mu \psi_\nu^I \Big)
\non[1ex]
&& + e\Big( \ft18 (\Bchi \chi) (\Bchi \chi) - 
 \ft1{96} \Bchi 
\Gg^\mu \GG^{IJ}\chi\,\Bchi \Gg_\mu \GG^{IJ}\chi \Big)
\;.
\ea
The invariance is most conveniently checked in 1.5 order
formalism, with the torsion
\be
T_{\mu\nu}{}^\rho = 
\ft12 \I \Bpsi{}^K_\mu \Gg^\rho \psi_\nu^K
+ \ft14 \I\,  \Bchi^\dA\, \Gg_{\mu\nu}{}^\rho\chi^\dA
\;.
\ee

A central role in our construction is played by the on-shell duality
between scalar fields and vector fields in three dimensions, which we
shall now discuss. The scalar field equation induced by \Ref{Lag} is
given by
\ba
\lefteqn{D_\mu 
\left(e\,\big( P^{\mu A} - \Bpsi^I_\nu \Gg^\mu \Gg^\nu 
       \chi^\dA \GG^I_{A\dA}\big)\right)  ~=} 
\non[.5ex]
&=& \ft12\, \Ge^{\mu\nu\rho}\Bpsi_\mu^I \psi_\nu^J 
    \GG^{IJ}_{AB} P_\rho^B
+\ft18\, \I e \,\Bchi \Gg^\mu \GG^{IJ} \chi\,
 \GG^{IJ}_{AB} P^B_\mu \;,
\ea
Upon use of the Rarita-Schwinger and Dirac equations for $\psi_\mu^I$
and $\chi^\dA$, respectively, this equation may be rewritten in the
form
\be
\dd^\mu \left(e\,\cJ_\mu{}^\cM\right) ~=~ 0 \;,
\la{concur}
\ee 
where $\cJ_\mu{}^\cM$ is the conserved Noether current associated
with the rigid $E_{8(8)}$ symmetry \cite{Nico91b}:
\ba
e\cJ^\mu{}^\cM &=& 2\VV{\cM}{B}\widehat{P}^{\,\mu B}
-\ft{\I}2\VV{\cM}{IJ}\,
{\Bchi}\Gg^\mu\GG^{IJ}\chi \non
&&{}-2e^{-1}\Ge^{\mu\nu\rho}
\left(\VV{\cM}{IJ}\,{\Bpsi}{}^I_\nu\psi^J_\rho
-\I\,\GG^I_{A\dA}\VV{\cM}{A}\,{\Bpsi}{}^I_\nu
\Gg^{\vphantom{A}}_\rho\chi^\dA \right) \;.
\la{J}
\ea
In writing this expression we have made use of the equivalence of the
fundamental and adjoint representations of $E_{8(8)}$ which yields the
relation (see also App.~A)
\ben
\VV{\cM}{\cA} := \ft1{60} {\rm Tr} \, \big( t^\cM
\, \cV \,t_\cA \,\cV^{-1} \big) \;.
\een
The existence of the conserved current \Ref{J} allows us to introduce
248 abelian vector fields ${B_\mu}^\cM$ (with index $\CM=1, \dots ,
248$), via
\be
\Ge^{\mu\nu\rho}\,B_{\nu\rho}{}^\cM ~=~ e\cJ^{\mu\cM} \;,
\la{dual1}
\ee
where $B_{\mu\nu}{}^\cM := \dd_\mu {B_\nu}^\cM\!-\!\dd_\nu
{B_\mu}^\cM$ denotes the abelian field strength. This equation defines
the vector fields up to the $\big[ U(1)\big]^{248}$ gauge transformations
\be
B_\mu{}^\cM\ra B_\mu{}^\cM + \dd_\mu\Lambda^\cM\;.
\ee
In accordance with \Ref{GHV} these vector fields transform in the
adjoint representation of rigid $\EE$ and are singlets under local
$SO(16)$. The supersymmetry transformations of the vector fields have
not been given previously; they follow by ``$E_{8(8)}$
covariantization'' of the supersymmetry variations of the 36 vector
fields obtained by direct dimensional reduction of $D\equ 11$
supergravity to three dimensions~\cite{KoNiSa00}
\be\la{susyv}
\Gd B_\mu{}^\cM ~=~ -2 \,\VV{\cM}{IJ}\,{\Beps}^I\psi^J_\mu
+\I \GG^I_{A\dA}\,\VV{\cM}{A}\,{\Beps}^I\Gg_{\mu}\chi^\dA \;.
\ee
For consistency, this transformation must be compatible with the
duality relation \Ref{dual1}. To check this, it is convenient to
rewrite the latter in terms of the supercovariant field strength
\ba
\widehat{B}_{\mu\nu}{}^\cM &:=& B_{\mu\nu}{}^\cM + 
 2\,\VV{\cM}{IJ}\,{\Bpsi}{}^I_\mu\psi^J_\nu
-2\I\,\GG^I_{A\dA}\VV{\cM}{A}\,{\Bpsi}{}^I_{[\mu}
\Gg^{\vphantom{A}}_{\nu]}\chi^\dA
\;, 
\la{supercovariant}
\nn
\ea
whose supercovariance is straightforwardly verified from
\Ref{susyv}. The duality relation \Ref{dual1} then takes the
following supercovariant form
\be\la{duality}
\Ge^{\mu\nu\rho} \widehat{B}_{\nu\rho}{}^\cM =
2e\,\VV{\cM}{A}\widehat{P}^{\,\mu A}
-\ft{\I}2\,e\VV{\cM}{IJ}\,
{\Bchi}\Gg^\mu\GG^{IJ}\chi \;.
\ee

Equation \Ref{duality} consistently defines the dual vector fields as
nonlocal and nonlinear functions of the original 248 scalar fields
(including the 120 gauge degrees of freedom associated with local 
$SO(16)$), provided the latter obey their equations of motion. 
We emphasize that in this way we can actually introduce as many 
vector fields as there are scalar fields, whereas the direct 
dimensional reduction of $D\!=\!11$ supergravity to three dimensions 
produces only 36 vector fields. The  ``$E_{8(8)}$ covariantization''
alluded to above simply consists in extending the relevant formulas from
these 36 vectors to the full set of ${\rm dim} \, G_0 \leq 248$ vector 
fields in a way that respects the $E_{8(8)}$ structure of the theory.
In the ungauged theory the vector fields have been introduced merely
on-shell; there is no Lagrangian formulation that would comprise the
scalar fields as well as their dual vector fields. However, we shall 
see that the gauged theory provides a natural off-shell framework
which accommodates both the scalars and their dual vectors.

{}From \Ref{duality} we can also extract the equation of motion of the
dual vectors: acting on both sides with $\Ge_{\rho\mu\nu} \dd^\nu$ and
making use of the integrability relations \Ref{int}, we obtain
\ba
\dd_\nu B^{\mu \nu}{}^\cM &=& 
-\ft12\,e^{-1}\,\Ge^{\mu\nu\rho}\, 
\VV{\cM}{IJ}\, Q^{IJ}_{\nu\rho} ~+~ \mbox{fermionic terms} \;.
\la{eqmB}
\ea
Also the fermionic terms still depend on the original scalar
fields. This is obvious from the fact that we need the scalar field
matrix $\cV$ to convert the $SO(16)$ indices on the fermions into the
$E_{8(8)}$ indices appropriate for the l.h.s. of this equation.  (Let
us note already here that in the gauged theory, the r.h.s.\ of this
equation will acquire additional contributions containing
$B_{\mu\nu}{}^\cM$ in order of the coupling constant).  We recognize
an important difference between the ``dual formulations'' of the
theory: whereas the vectors disappear completely in the standard
formulation of the theory, the vector equations of motion in general
still depend on the dual scalar fields. It is only under very special
circumstances, and for special subsets of the 248 vector fields, that
one can completely eliminate the associated dual scalars. This is
obviously the case for the version obtained by direct reduction of
$D=11$ supergravity to three dimensions where only 92 bosonic degrees
of freedom appear as scalar fields while 36 physical degrees of
freedom appear as vector fields.  As shown in \cite{CJLP98}, the
latter are associated with the 36-dimensional maximal nilpotent
commuting subalgebra of $E_{8(8)}$, but there are further intermediate
possibilities.

To conclude this section, we recall that the three dimensional 
Einstein-Hilbert term can be rewritten in Chern-Simons form as
\be
-\ft14 e R= \ft14\,\Ge^{\mu\nu\rho}\,e_\mu{}^a\, F_{\nu\rho\,a} \;,
\ee
by means of the dual spin connection
\ben
A^a_{\mu}=-\ft12\Ge^{abc}\Go_{\mu\,bc} \;,
\een
with field strength $F^a_{\mu\nu}=
2\dd_{[\mu}A^a_{\nu]}+\Ge^a{}_{bc}\,A^b_{\mu}A^c_{\nu}$. When gauging
the theory the Minkowski background space-time will be deformed to an
AdS$_3$ spacetime characterized by
\be\la{Einstein}
R_{\mu\nu} = 2m^2 g_{\mu\nu} \;,
\ee
with (negative) cosmological constant $\Lambda = -2m^2$. The
Lorentz-covariant derivative is accordingly modified to an 
AdS$_3$ covariant derivative
\be
{\cal D}^\pm_\mu := \partial_\mu + 
\ft12 \I \Gg_a (A_\mu{}^a \pm m e_\mu{}^a) \;,
\la{covAdS}
\ee
with commutator
\ben
[{\cal D}^\pm_\mu ,  {\cal D}^\pm_\nu ] =
\ft12 \I \Gg_a ( F_{\mu\nu}{}^a + m^2 \Ge^{abc} e_{\mu b} e_{\nu c} )
\;.
\een
We will return to these formulas when discussing the conditions
for $(n_L,n_R)$ supersymmetry in AdS$_3$ in Chapter~\ref{CHgs}.

\mathon
\section{Gauged $N\equ 16$ supergravity}
\mathoff
\la{CHgauge}

The Lagrangian \Ref{Lag} is invariant under rigid $\EE$ and local
$SO(16)$. To gauge the theory, we now select a subgroup $G_0\subset\EE$ 
which will be promoted to a local symmetry. The resulting theory will 
then be invariant under local $G_0 \times SO(16)$, such that 
\Ref{GHV} is replaced by
\be
\cV (x)\; \longrightarrow \; g_0 (x)\, \cV(x) \,h^{-1}(x)\;, \qquad
  g_0 (x) \in G_0 \; , \; h(x) \in SO(16) \;,
\label{GHV1}
\ee
However, it should be kept in mind that the local symmetries are 
realized in different ways: as before, the local $SO(16)$ is 
realized in terms of ``composite'' gauge connections, whereas 
the gauge fields associated with the local $G_0$ symmetry are 
independent fields to begin with. Restricting to semisimple 
subgroups, $G_0$ is properly characterized by means of its embedding 
tensor $\Theta_{\cM\cN}$ which is the restriction of the
Cartan-Killing form $\eta_{\cM\cN}$ onto the associated algebra
$\mfg_0$. The embedding tensor will have the form
\be
\Theta_{\cM\cN} = \sum_j \varepsilon_j \, \eta^{(j)}_{\cM\cN} \;,
\la{emb}
\ee
where $\eta^{\cM\cN} \eta^{(j)}_{\cN\cK}$ project onto the simple
subfactors of $G_0$, and the numbers $\varepsilon_j$ correspond to the
relative coupling strengths. It will turn out that these coefficients
are completely fixed by group theory, so there is only one overall gauge
coupling constant $g$. Owing to the symmetry of the projectors
$\eta^{(j)}$ the embedding tensor is always symmetric:
\be
\Theta_{\cM\cN} = \Theta_{\cN\cM} \;.
\la{symTH}
\ee
As discussed in the introduction we introduce a subset of $\dimg={\rm
dim~} G_0$ vector fields, obtained from \Ref{duality} by projection
with $\Theta_{\cM\cN}$. For these we introduce special labels
$m,n,\dots$, with the short hand notation
\be\la{m}
{B_\mu}^m t_m \equiv {B_\mu}^\cM \,\Theta_{\cM\cN} \, t^\cN \;,
\qquad {\rm etc.}
\ee
Note that we do not make any assumption about $G_0$ at this point; in 
particular, our ansatz allows for compact as well as noncompact gauge
groups. The possible choices for $G_0$ will be determined in 
Chapter~\ref{CHG0}. 

The first step is the covariantization of derivatives in
\Ref{VdV} according to
\be
\la{gauging}
\cV^{-1}\CD_{\!\mu} \cV ~\equiv~
\cV^{-1}\dd_\mu \cV 
+ g\,B_\mu{}^m \, \cV^{-1} t_m  \cV ~\equiv~ 
\CP_\mu^AY^A +\ft12 \CQ_\mu^{IJ}X^{IJ} \;,
\ee
with gauge coupling constant $g$. The non-abelian field strength reads
\ba\la{FS}
\CB_{\mu\nu}{}^m &:=& \dd_\mu\,B_{\nu}{}^m - \dd_\nu\,B_{\mu}{}^m
+ g\,f^{m}{}_{np}\,B_{\mu}{}^n B_{\nu}{}^p \;.
\ea
The integrability relations \Ref{int} are modified to
\ba
\CQ_{\mu\nu}^{IJ}+\ft12\,\GG^{IJ}_{AB}\,\CP^A_\mu \CP^B_\nu &=&
g\,\CB_{\mu\nu}{}^m \,\GTh_{mn} \VV{n}{IJ} \;, \non[1ex]
2 D^{\vphantom I}_{[\mu}\CP_{\nu]}^A &=& 
 g\,\CB_{\mu\nu}{}^m \,\GTh_{mn} \VV{n}{A}
\;.
\la{m2}
\ea
With the hidden $g$ dependent extra terms in the definition of
the currents in \Ref{gauging}, their supersymmetry variations become
\ba
\Gd \CQ_\mu^{IJ} &=& \ft12\,
(\GG^{IJ}\GG^K)_{A\dot A} \CP_\mu^A \,\Bchi^{\dot A} \Ge^K  \, 
+ \, 
    g (\Gd B_\mu{}^m) \,\GTh_{mn} \VV{n}{IJ} \;, \non[1ex]
\Gd \CP_\mu^{A} &=& 
\GG^I_{A\dot A} \,D_\mu ( \Bchi^{\dot A} \Ge^I )
+ \,
    g (\Gd B_\mu{}^m)\, \GTh_{mn} \VV{n}{A} 
\;,
\la{m1}
\ea
with the variation of the vector fields given in \Ref{susyv}.

Both modifications violate the supersymmetry of the original
Lagrangian.  In order to restore local supersymmetry we follow the
standard Noether procedure as in \cite{deWNic82}, modifying both the
original Lagrangian as well as the transformation rules by
$g$-dependent terms. We will first state the results, and then explain
their derivation and comment on the special and novel features of our
construction.

The full Lagrangian can be represented in the form
\be
\CL = \CL^{(0)} + \CL^{(1)} + \CL^{(2)} + \CL^{(3)} \;,
\la{L0123}
\ee
where $\CL^{(0)}$ is just the original Lagrangian \Ref{Lag}, but
with the modified currents defined in \Ref{gauging}; thus $\CL^{(0)}$
and $\CL$ differ by terms of order $\cO(g)$. The contributions
$\CL^{(1)}$ and $\CL^{(2)}$ are likewise of order $g$ and describe the
Chern-Simons coupling of the vector fields and the Yukawa type
couplings between scalars and fermions, respectively:
\ba
\CL^{(1)} &=& 
-\ft14\,g\,\Ge^{\mu\nu\rho}\,B_\mu{}^m
\Big(\dd_\nu B_\rho\,{}_m
+\ft13\,g f_{mnp}\,B_\nu{}^n B_\rho{}^p \Big) 
\;,
\la{L1}
\ea
\ba
\CL^{(2)} &=& 
\ft12ge\,
A_{1}^{IJ}\;{\Bpsi}{}^I_{\mu}\,\Gg^{\mu\nu}\,\psi^{J}_{\nu} + 
{\I}ge\,A_{2}^{I\dA}\;
{\Bchi}{}^\dA\,\Gg^\mu\,\psi^I_{\mu}
\non
&&{}
+ \ft12ge\,
A_{3}^{\dA\dB}\;{\Bchi}{}^\dA\,\chi^{\dB}\;,
\la{L2}
\ea
where the tensors $A_{1,2,3}$ are functions of the scalar matrix $\cV$
which remain to be determined. At order $\cO(g^2)$, there is the
scalar field potential $W(\cV)$:
\ba\la{potential}
\CL^{(3)} &=& eW ~\equiv~
\ft18 \,g^2\,e\,\Big(
A_{1}^{IJ}A_{1}^{IJ}-\ft12\,A_{2}^{I\dA}A_{2}^{I\dA} \Big) \;.
\ea
Besides the extra $g$ dependent terms induced by the modified currents, 
the supersymmetry variations must be amended by the following
$\cO(g)$ terms:
\be
\Gd_g \psi^I_\mu = \I g\,{A}_{1}^{IJ}\,\Gg_{\mu} \Ge^J \;, \qquad
\Gd_g \chi^\dA =  g\,A_{2}^{I\dA}\,\Ge^I \;.  \la{ferm}
\ee
Of course, the above modifications of the Lagrangian and the supersymmetry
transformation rules have not been guessed ``out of the blue'', but 
at this point simply constitute an ansatz that has been written down
in analogy with known gauged supergravities, in particular the $N\equ 8$ 
theory of \cite{deWNic82}. The consistency of this ansatz must now
be established by explicit computation.

The $SO(16)$ tensors $A_{1,2,3}$ depending on the scalar fields $\cV$
introduce Yukawa-type couplings between the scalars and the fermions 
beyond the derivative couplings generated by \Ref{VdV}, as well as a 
potential for the scalar fields. As is evident from their definition, 
the tensors $A_1^{IJ}$ and $A_3^{\dA\dB}$ are symmetric in their respective
indices. Therefore, $A_1^{IJ}$ decomposes as ${1} + {135}$ under
$SO(16)$,\footnote{Here and in the following, representations of
$SO(16)$ are written with ordinary numbers, while representations of
$E_{8(8)}$ are given in boldface numbers.} viz.
\be
A_1^{IJ} = A_1^{(0)} \Gd^{IJ} + \TA_1^{IJ} \;,
\ee
with $\TA^{JJ} = 0$, while for $A_3^{\dA\dB}$ we have the decomposition
\be
A_3^{\dA\dB} = A_3^{(0)} \Gd^{\dA\dB} + \TA_3^{\dA\dB} \;,
\ee
where
\ben
\TA_3^{\dA\dB} = 
   \ft1{4!}  A^{(4)}_{3\, IJKL} \GG^{IJKL}_{\dA\dB} +
   \ft1{2\cdot 8!}  A^{(8)}_{3\, I_1...I_8} \GG^{I_1...I_8}_{\dA\dB} 
\;.
\een
Therefore $A_3$ can contain the representations ${1}+ {1820} + {6435}$. 
However, we will see that the ${6435}$ drops out. Due to the
occurrence of the $1820$ in this decomposition, the tensor $A_3$
cannot be expressed in terms of $A_{1,2}$ unlike for $D=4$ and $D=5$.
The independence of $A_3$ is a new feature of the $D=3$ gauged theory.

Several restrictions on the tensors $A_{1,2,3}$ can already 
be derived by imposing closure of the supersymmetry algebra on 
various fields at order $\cO(g)$. Computing the commutator on the
dreibein field we obtain an extra Lorentz rotation with parameter
\be
\Lambda_{\Ga\Gb} = 2g A_1^{IJ}\, \Beps_1^I \Gg_{\Ga\Gb} \Ge_2^J
\;,
\ee
while evaluation of the commutator on the vector fields and the
scalar field matrix $\cV$ yields an extra gauge transformation
with parameter
\be
\GL^m = 
2\,\VV{m}{IJ}\,{\Beps}{}^I_1\Ge^J_2 
+\I B_\mu{}^m\,{\Beps}{}^I_1\Gg^\mu \Ge^I_2 \;.
\ee
The latter induces a further $SO(16)$ rotation with parameter 
$\omega^{IJ} = g \GL_m {\cV^m}_{IJ}$ on $\cV$ (as well as the fermions
which transform under $SO(16)$). For the derivation of this result
we need the relations
\ba
\VV{m}{A}\,\GG^{(I}_{A\dA}\,A_2^{J)\dA} &=&
\VV{m}{IK}\,A_1^{JK} + \VV{m}{JK}\,A_1^{IK} \;,
\la{s1}\\
\GG^{[I}_{A\dA}\,A_2^{J]\dA}&=& 
\VV{\cC}{IJ}\GTh_{\cC\cD}\VV{\cD}{A}
\;,
\la{l1}
\ea
which give the first restrictions on the tensors $A_{1,2,3}$.
A peculiarity is that the closure of the superalgebra on $B_\mu{}^m$
requires use of the duality equation, whereas the equations of
motion are not needed to check closure on the remaining bosonic fields.

Tracing \Ref{s1} over the indices $I$ and $J$ and using 
the symmetry of $A_1^{IJ}$ we immediately obtain
\be\la{1920}
\GG^I_{A\dA} A_2^{I\dA} = 0 \;.
\ee
The tensor $A_2^{I\dA}$ thus transforms as the $\overline{1920}$
(traceless vector spinor) representation of $SO(16)$.

To state the restrictions imposed on these tensors by the 
requirement of local supersymmetry more concisely, we now 
define the $T$-tensor
\be\la{defT}
\TT{\cA}{\cB} := \VV{\cM}{\cA} \VV{\cN}{\cB} \,\Theta_{\cM\cN} \;.
\ee
Clearly $T_{\cA|\cB}= T_{\cB|\cA}$ by the symmetry of $\Theta$.
Unlike the cubic expressions in \cite{deWNic82} and \cite{GuRoWa86}, 
however, the $T$-tensor is quadratic in $\cV$ due to the equivalence
of the fundamental and adjoint representations for $E_{8(8)}$, see
\Ref{adjoint}. The tensors $A_{1,2,3}$ must be expressible in terms 
of $T$ if the theory can be consistently gauged. The detailed 
properties of the $T$-tensor will be the subject of the following chapter.

Let us next consider the consistency conditions for local supersymmetry
of \Ref{L0123} step by step. All cancellations that are
$G_0$-covariantizations of the corresponding terms in the ungauged
theory will work as before, and for this reason we need only discuss
those variations which have no counterpart in the ungauged theory.
Variation of $\CL^{(1)}$ produces only the contribution
\ben
\Gd \CL^{(1)} =
- \ft14 g\Ge^{\mu\nu\rho} \Gd B_\mu{}^m \CB_{\nu\rho m}
\;,
\een
because the CS term depends on no other fields but $B_\mu{}^m$.
Inserting \Ref{susyv} the above variation can be seen to cancel
against the extra terms in the variation of $\CL^{(0)}$ arising
in the integrability conditions, cf. \Ref{m2}

A second set of $g$-dependent terms is obtained by varying $B_\mu{}^m$ 
in $\CQ_\mu$ and $\CP_\mu$, cf.~\Ref{m1}. Expressing the result by means
of the $T$-tensor, we obtain
\ba
\lefteqn{g \left( 2\, \TT{IJ}{KL} \, \Beps^I \psi_\mu^J  -
\I\, \TT{KL}{A} \,\GG^I_{A\dB} \Beps^I \Gg_\mu \chi^\dB \right) 
\left( \Bpsi_\nu^K\Gg^{\mu\nu\rho}\,\psi_\rho^L + 
            \ft{\I}4\Bchi\Gg^\mu \GG^{KL} \chi \right)}     
\non[.5ex]
\lefteqn{ -\; g\left( \TT{A}{KL}\, \Beps^K\psi_\mu^L -
\ft12\I\, \TT{A}{B}\,\GG^K_{B\dB} \Beps^K \Gg_\mu \chi^\dB \right) 
    \left( \CP^{\mu A} - 
\Bchi^\dA \Gg^\nu \Gg^\mu \psi_\nu^I \,\GG^I_{A\dA} \right)
\;.}
\nn
\ea
These terms combine with the variations of the fermionic fields from
$\CL^{(2)}$ and the new variations \Ref{ferm} in
$\CL^{(0)}$. Consideration of the $\Ge\psi\CP$ and $\Ge\chi\CP$ terms
now reproduces \Ref{l1}, but in addition requires the differential
relations
\ba
\CD_\mu A_{1}^{IJ} &=& \CP_\mu{}^A\,\GG^{(I}_{A\dA}\,A_2^{J)\dA} \;, 
\non
\CD_\mu A_{2}^{I\dA} &=& 
\ft12\,\CP_\mu{}^A\,
\Big(\GG^I_{A\dB}\,A_3^{\dA\dB} + \GG^J_{A\dA}\,A_1^{IJ} \Big) 
     -\ft12\,\CP_\mu{}^A\, \GG^I_{B\dA}\, \TT{A}{B} \;.
\la{id2}
\ea
Multiplying the second relation by $\GG^I_{A\dA}$ and invoking
\Ref{1920} yields
\be\la{id1}
\TT{A}{B} = \big( A_1^{(0)} + A_3^{(0)} \big) \, \Gd_{AB} + 
\ft1{16} \GG^I_{A\dA} \,\TA_3^{\dA\dB} \,\GG^I_{\dB B} \;.
\ee
Since $\GG^I \GG^{(8)} \GG^I = 0$ there is no $\overline{6435}$ of
$SO(16)$ in $T_{A|B}$. However, the argument does not yet suffice 
to rule out such a contribution in $A_3$.

As in \cite{deWNic82}, the supersymmetry variation of the tensors 
$A_{1,2}$ is obtained from \Ref{id2} by replacing $\CP_\mu^A$ by 
$\GG^I_{A\dA}\Beps^I\chi^\dA$:
\ba
\Gd \TA_{1}^{IJ} &=& 
\GG^K_{A\dB} \Beps^K\chi^\dB\,\GG^{(I}_{A\dA}\,A_2^{J)\dA} \;, 
\non
\Gd A_{2}^{I\dA} &=& 
\ft12\,\GG^K_{A\dB} \Beps^K\chi^\dB
\Big(\GG^I_{A\dC}\,A_3^{\dA\dC} + \GG^J_{A\dA}\,A_1^{IJ}  
-\GG^I_{B\dA}\, \TT{A}{B} \Big) 
\;.
\la{varA}
\ea
The tracelessness of $A_2^{I\dA}$ in \Ref{1920} in conjunction
with \Ref{id2} also implies that $A_1^{(0)}$ and $A_3^{(0)}$ 
are constant. This is consistent with the fact that the trace 
parts drop out from the above variations. Observe that the 
supersymmetry variation of $A_3$ does not yet enter at 
this point as it appears only at cubic order in the fermions.

At $\cO(g^2)$ we get two quadratic identities. The first multiplies 
the $g^2 \psi\Ge$ variations and is straightforwardly obtained
\ba
 A_1^{IK}A_1^{KJ} - \ft12\,A_2^{I\dA}A_2^{J\dA}
&=& \ft1{16}\,\Gd^{IJ}\,
\Big(
A_1^{KL}A_1^{KL} - \ft12\,A_2^{K\dA}A_2^{K\dA}
\Big) \;.
\la{qu1}
\ea
The second comes from the $g^2 \chi\Ge$ variations: performing
the $\cO(g)$ variations in $\CL^{(2)}$ we obtain
\ben
\Gd_g \CL^{(2)} = 
g^2 e \Bchi^\dA \Ge^I \big( - 3 A_1^{IJ} A_2^{I\dA}
   + A_3^{\dA\dB} A_2^{I\dB} \big) 
\;.
\een
Varying $A_{1,2}$ in the potential, on the other hand, and making use
of the above formulas \Ref{varA} together with \Ref{1920}, we arrive
at:
\ben
\Bchi^\dA \Ge^K (\GG^K \GG^I)_{\dA\dB} \left( \ft3{16} \TA_1^{IJ}
   A_2^{J\dB} - \ft1{16} \TA_3^{\dB\dot C} A_2^{I\dot C} \right)
\;.
\een
By the tracelessness of $A_2^{I\dA}$ we can drop the tildes in
this expression, and thus obtain the second relation
\ba
3A_1^{IJ}A_2^{J\dA}\!-\!A_2^{I\dB}A_3^{\dA\dB}
\!
&=&\! \ft1{16}\,(\GG^I\GG^J)_{\dA\dB}\,
\Big(3A_1^{JK}A_2^{K\dB}\!-\!A_2^{J\dC}A_3^{\dB\dC}\Big) 
\;,
\la{qu2}
\ea
which must be satisfied for local supersymmetry to hold.
 
Thus, at linear order in the fermions, supersymmetry requires the
tensors $A_{1,2,3}$ to satisfy the identities \Ref{s1}, \Ref{l1}, and
\Ref{id2}--\Ref{qu2}. However, these do not yet constitute a complete
set of restrictions. In marked contrast to the $D\!\geq\!4$ gauged
supergravities, we get further and independent conditions at cubic
order in the fermions. This special feature is again related
to the algebraic independence of the third tensor $A_3$. 
Although the necessary calculations are quite tedious, we here
refrain from giving details and simply state the results, as the
relevant Fierz technology is (or should be) standard by
now. Interested readers may find many relevant formulas in
\cite{MarSch83}. 

The analysis of the $(\Bpsi\psi)(\Bpsi\Ge)$ terms gives
\be
\TT{IJ}{KL} = 2 \Gd\oversym{_{\vphantom{1}}^{I[K} A_1^{L]J}\,} 
+ \TT{[IJ}{KL]}
\;.
\la{id4}
\ee
The structure of the r.h.s. of this equation thus restricts
$T_{IJ|KL}$ to the $SO(16)$ components ${1}$, ${135}$ and ${1820}$. 
Demanding the cancellation of $(\Bchi\chi)(\Bpsi\Ge)$ terms yields 
three more constraints:  
\ba
A_3^{(0)} + 2 A_1^{(0)} &=& 0 \;,\non
A^{(8)}_{3\, I_1...I_8} &=& 0 \;,\non
\TT{[IJ}{KL]} &=& 2 \TA^{(4)}_{3\,IJKL} \;,
\la{A3}
\ea
such that with \Ref{l1}, \Ref{id1}, and \Ref{id4} the $T$-tensor
\Ref{defT} may entirely be expressed in terms of the tensors
$A_{1,2,3}$: 
\ba
\TT{IJ}{KL} &=& 
2\,\Gd^{IJ}_{KL}\,A_1^{(0)} +
2\, \Gd\oversym{_{\vphantom{1}}^{I[K} \TA_1^{L]J}\,} 
+\,  2\, \TA^{(4)}_{3\,IJKL} \;,\non
\TT{IJ}{A}&=& \GG^{[I}_{A\dA}\,A_2^{J]\dA}
\;,\non
\TT{A}{B} &=& -A_1^{(0)} \Gd_{AB} + 
\ft1{2\cdot 4!}\, \GG^{IJKL}_{AB}\, \TA^{(4)}_{3\,IJKL} 
\;.
\la{l2}
\ea
In particular, the two singlets and the two $1820$ representations in
$T_{IJ|KL}$ and $T_{A|B}$ coincide. Finally, the analysis of the
$(\Bchi \chi)(\Bchi\Ge)$ terms yields
\be
\delta A_{3\,IJKL}^{(4)} = -\ft12\, 
\Beps^M\chi^\dA \left(\GG^M \GG^{[IJK}\right)_{\dA\dB} A_2^{L]\dB}
\;.
\la{d2}
\ee
In order to derive this condition and to prove the vanishing of the
$(\Bchi \chi)(\Bchi\Ge)$ terms, one needs the additional Fierz
identity, which cannot be derived from the relations given in 
the Appendix of \cite{MarSch83}:
\ba
\lefteqn{(\Bchi\GG^{KLMN}\chi)\,
(\Bchi^\dA\Ge^I) \,(\GG^I\GG^{KLM})_{\dA\dB}A_2^{N\dB} ~=} \non[.7ex]
&=&{} 
36\,(\Bchi\Gg_\mu\GG^{IJ}\chi)\, 
(\Bchi^\dA\Gg^\mu\Ge^I)\, A_2^{J\dA}
-4\,(\Bchi\Gg_\mu\GG^{KL}\chi)\, 
(\Bchi^\dA\Gg^\mu\Ge^I)\, \GG^{KL}_{\dA\dB} A_2^{I\dB}
\non[.1ex]
&&{}
+
48\,(\Bchi\chi)\, 
(\Bchi^\dA\Ge^I)\,  A_2^{I\dA} 
-12\,(\Bchi\Gg_\mu\GG^{KL}\chi)\, 
(\Bchi^\dA\Gg^\mu\Ge^I)\, \GG^{IK}_{\dA\dB}A_2^{L\dB} \;,
\nn
\ea
The tracelessness of $A_2^{I\dA}$ is again crucial in obtaining 
this result. 

Let us summarize our findings. The complete set of consistency
conditions ensuring supersymmetry of the gauged Lagrangian \Ref{L0123}
is given by the linear relations \Ref{l2}, the differential identities
\Ref{id2}, \Ref{d2}, the relation \Ref{s1}, and the quadratic
identities \Ref{qu1}, \Ref{qu2}.  The tensors $A_{1,2,3}$ can contain
only the $SO(16)$ representations ${1}, {135}, {1820}$ and
$\overline{1920}$. Equations \Ref{l2} show that likewise the
$T$-tensor may contain only these representations. The remarkable fact
-- which eventually allows the resolution of all identities --
is that these $SO(16)$ representations combine into representations 
of $E_{8(8)}$. More specifically, we have
\be
{135}\!+\!{1820}\!+\!\overline{1920} ~=~ {\bf 3875}\;,
\la{break}
\ee
while the first relation from \Ref{A3} ensures that the two $SO(16)$
singlets originate from one singlet of $E_{8(8)}$, such that the full
$E_{8(8)}$ content of the tensors $A_{1,2,3}$ is contained in the
$E_{8(8)}$ representations $\bf{1} + \bf{3875}$.  Apart from the
occurrence of an extra singlet, this fusion of tensors into
representations of the hidden global $E_{d(d)}$ takes place already in
dimensions $D\equ 4$ and $D\!=5$, where the Yukawa couplings are given
by tensors transforming in the ${\bf 912}$ of $E_{7(7)}$
\cite{deWNic84} and in the ${\bf 351}$ of $E_{6(6)}$ \cite{GuRoWa86},
respectively.  We shall come back to this point in the next chapter.

Perhaps the most unexpected feature of our construction is the fact
that the vector fields appear via a CS term \Ref{L1} in order $g$,
rather than the standard Yang-Mills term. This has no analog in higher
dimensions, where the vector fields appear already in the ungauged
theory via an abelian kinetic term. In hindsight this coupling of the
vector fields turns out to be the only consistent way to bring in the
dual vector fields without introducing new propagating degrees of
freedom, and thereby to preserve the balance of bosonic and fermionic
physical degrees of freedom.

The emergence of non-abelian CS terms in the maximally supersymmetric 
theories naturally leads to a {\em non-abelian extension of the 
duality relation}
\Ref{duality}
\be\la{NAduality}
\Ge^{\mu\nu\rho} \widehat{\CB}_{\mu\nu}{}^m =
2\,e\VV{m}{A}\,\widehat{\CP}^{\,\rho A}
-\ft{\I}2\,e\VV{m}{IJ}\,
{\Bchi}\Gg^\rho\GG^{IJ}\,\chi \;,
\ee
which consistently reduces to \Ref{duality} in the limit $g\rightarrow
0$. However, in this limit, the vector fields drop from the Lagrangian
such that the duality relation \Ref{duality} no longer follows
from a variational principle in the ungauged theory but rather must be
imposed by hand. This can be viewed as a very mild form of the
gauge discontinuity encountered for gauged supergravities in odd 
dimensions \cite{PePivN84,PePivN85,GuRoWa86}. In contrast to those
models however, the Lagrangian \Ref{L0123} has a perfectly
smooth limit as $g\rightarrow0$.

Because of the explicit appearance of the gauge fields on the r.h.s.\
of the non-abelian duality relation it is no longer possible to trade
the vector fields for scalar fields and thereby eliminate them, unlike
in \cite{CJLP98}.  Vice versa, the explicit appearance of the scalar
fields in the potential of \Ref{L0123} also excludes the possibility to
eliminate some of these fields by replacing them by vector fields. In
contrast to the ungauged theory which allows for different equivalent
formulations related by duality, the gauged theory apparently comes in
a unique form which requires the maximal number of scalar fields
together with the dual vectors corresponding to the gauge group $G_0$.

Note that unlike in \Ref{duality}, the nonabelian duality relation
\Ref{NAduality} may be imposed only for those vector fields which
belong to the gauge group $G_0$. Having gauged the theory, we 
can no longer introduce additional vector fields as was the case 
for the ungauged theory. This is because additional vector fields 
transforming nontrivially under the gauge group $G_0$ would 
acquire mass terms in the gauged theory, entailing a mismatch
between bosonic and fermionic degrees of freedom. As a consequence,
\Ref{NAduality} does not imply the full set of bosonic equations
of motion, but just their projection onto the subgroup $G_0$. However,
just as in \Ref{eqmB} we may deduce the equations of motion for 
the vector fields from \Ref{NAduality} by acting on both
sides with $\Ge_{\rho\mu\nu} \CD^\nu$ and making use of \Ref{m2}:
\ba
\CD_\nu \CB^{\mu \nu}{}^m &=& 
\ft12\,ge^{-1}\Ge^{\mu\nu\rho}\,
\left(\VV{m}{A}\VV{n}{A}+\VV{m}{IJ}\VV{n}{IJ} \right)\,
\GTh_{nk}\,\CB_{\nu\rho}{}^k \non
&&{}-\ft12\,e^{-1}\Ge^{\mu\nu\rho}\,
\VV{m}{IJ}\, \CQ^{IJ}_{\nu\rho}
\;\;+ \mbox{fermionic terms} \non[2ex]
&=& g\left( \VV{m}{B}\TT{B}{A} + \VV{m}{IJ}\TT{IJ}{A} \right) 
\CP^\mu{}^A 
-\ft12\,e^{-1}\Ge^{\mu\nu\rho}\,
\VV{m}{IJ}\, \CQ^{IJ}_{\nu\rho}
\non[1ex]
&&{}+ \mbox{fermionic terms}
\;.
\la{eqmBgauge}
\ea
%


\mathon
\section{$T$-identities}
\mathoff
\la{CHT}

In the foregoing chapter we have derived the consistency conditions
which must be satisfied by the tensors $A_{1,2,3}$ and the $T$-tensor
in order to ensure the full supersymmetry of the gauged action
\Ref{L0123}. It remains to show that these conditions admit nontrivial
solutions $A_{1,2,3}(\cV)$. This will single out the possible gauge
groups $G_0\subset\EE$.  Recall that in the three dimensional model the
choice of gauge group is less restricted than in higher dimensions
where the gauge group $G_0\subset G$ is essentially determined by the
fact that a maximal subset of the vector fields of the theory must
transform in its adjoint representation.

Up to this point, we have made no assumptions on the gauge group
$G_0\subset E_{8(8)}$, which is characterized by its embedding
tensor $\GTh_{\cA\cB}$, cf.~\Ref{emb}. We will now show that all the
consistency conditions derived in the previous section may be encoded
into a single algebraic equation for the embedding tensor.

According to \Ref{symTH}, $\GTh_{\cA\cB}$ transforms in the symmetric
tensor product
\be\la{sym248}
({\bf 248}\times{\bf 248})_{\rm sym} = 
{\bf 1}+{\bf 3875}+{\bf 27000} \;.
\ee
The explicit projectors of this decomposition have been computed in 
\cite{KoNiSa99} 
\ba 
({\mathbb P}_{\bf 1})_{\cM\cN}{}^{\cK\cL}     &=& 
\ft{1}{248}\,\eta_{\cM\cN} \,\eta^{\cK\cL}
\;,\non
({\mathbb P}_{\bf 3875})_{\cM\cN}{}^{\cK\cL}  &=& 
\ft{1}{7}\,  \Gd_{(\cM}^{\hphantom{(}\cK} \Gd_{\cN)}^{\cL} 
 -\ft{1}{56}\, \eta_{\cM\cN} \,\eta^{\cK\cL}
 -\ft{1}{14}\, f^\cP{}_\cM{}^{(\cK} f_{\cP\cN}{}^{\cL)}
\;,\non
({\mathbb P}_{\bf 27000})_{\cM\cN}{}^{\cK\cL} &=& 
\ft{6}{7}\, \Gd_{(\cM}^{\hphantom{(}\cK} \Gd_{\cN)}^{\cL} 
+\ft{3}{217}\,\eta_{\cM\cN} \,\eta^{\cK\cL}
+\ft{1}{14}\, f^\cP{}_\cM{}^{(\cK} f_{\cP\cN}{}^{\cL)}\;.
\la{projind}
\ea
Accordingly, $\GTh_{\cM\cN}$ may be decomposed as
\be
\GTh_{\cM\cN} = \Gth\,\eta_{\cM\cN}
+ \GTh_{\cM\cN}^{\bf 3875} + \GTh_{\cM\cN}^{\bf 27000} \;,
\la{TH123}
\ee
with
\ben
\GTh_{\cM\cN}^{\bf 3875} = 
({\mathbb P}_{\bf 3875})_{\cM\cN}{}^{\cK\cL} \,\GTh_{\cK\cL} \;,
\quad
\GTh_{\cM\cN}^{\bf 27000} = 
({\mathbb P}_{\bf 27000})_{\cM\cN}{}^{\cK\cL} \,\GTh_{\cK\cL} \;.
\een
The $T$-tensor as it has been defined in \Ref{defT} is given by a
rotation of $\GTh_{\cM\cN}$ by the matrix $\cV$. It may likewise 
be decomposed
\be\la{T}
\TT{\cA}{\cB} ~=~
\TTd{1}{\cA}{\cB} + \TTd{3875}{\cA}{\cB} + \TTd{27000}{\cA}{\cB} \;,
\ee
with
\ben
\TTd{3875}{\cA}{\cB}~=~ 
({\mathbb P}_{\bf 3875})_{\cA\cB}{}^{\cC\cD}\; \TT{\cC}{\cD}
~=~\VV{\cM}{\cA}\VV{\cN}{\cB}\,\GTh_{\cM\cN}^{\bf 3875}
\;,\qquad \mbox{etc.}
\een
where the second equality is due to invariance of the projectors
under $\EE$. Analogous tensors have been defined in \cite{deWNic82} 
and \cite{GuRoWa86} for the maximally gauged models in $D\equ 4$ and
$D\equ 5$, respectively. Unlike those $T$-tensors, however, the
$T$-tensor here is quadratic in $\cV$, as already emphasized before.

\subsection{The constraint for the embedding tensor}

We have seen that supersymmetry of the gauged Lagrangian in particular
implies the set of relations \Ref{l2} for the $T$-tensor. As discussed
above, these relations show that $T$ may only contain the $SO(16)$
representations contained in the ${\bf 1}\!+\!{\bf 3875}$ of $\EE$. It
follows that equations \Ref{l2} can be solved for $A_{1,2,3}$ if and
only if
\be\la{crit}
\TTd{27000}{\cA}{\cB}= 0 \quad \Longleftrightarrow \quad
\GTh_{\cA\cB}^{\bf 27000}= 0\;.
\ee
This is a set of linear algebraic equations for the
embedding tensor $\GTh_{\cA\cB}$. We stress once more the
remarkable fact that the equations \Ref{l2} combine into 
an $\EE$ covariant condition for the $T$-tensor which
makes it possible to translate these equations into a condition for
the constant tensor $\GTh$. In particular, each single equation from
\Ref{l2} yields an $SO(16)$ covariant restriction on the $T$-tensor
\Ref{defT} which already implies the full set of relations \Ref{l2},
if it is to be satisfied for all $\EE$ valued matrices $\cV$.

We shall show in the following sections that \Ref{crit} not only
reproduces the linear equations \Ref{l2} but indeed implies the
complete set of consistency conditions (including the differential 
and quadratic ones) identified in the last chapter\footnote{Let us
stress once more that in addition to \Ref{crit}, $\Theta$ must project 
onto a subgroup. If that condition is dropped, further solutions
to \Ref{crit} can be found, but the $T$-tensor would then fail to 
satisfy the quadratic identities of section 4.4.}.

\subsection{Linear identities}
Making use of the explicit form of the projectors \Ref{projind},
equation \Ref{crit} takes the form
\ba
\GTh_{IJ,KL}&=&
-\ft27\,\Gd\undersym{_{I[K}\,\GTh_{L]M,MJ}\,}
+ \GTh_{[IJ,KL]}
+\ft{16}{7}\,\Gth\,\Gd^{IJ}_{KL} \;,
\non[1ex]
\GTh_{IJ,A} &=& 
\ft17\,(\GG\undersym{_I\GG^L)_{AB}\,\GTh_{B,LJ}\,} 
\;,
\non[1ex]
\GTh_{A,B}
&=&
\ft1{96}\,\GG^{IJKL}_{AB}\,\GTh_{IJ,KL} 
+\Gth\,\Gd_{AB} \;,
\la{critex}
\ea
and likewise for $T$. These equations contain the complete set of
linear identities among different components of the $T$-tensor. Once
they are satisfied, the $T$-tensor may entirely be expressed in terms
of the tensors $A_{1,2,3}$ as found in \Ref{l2} above:
\ba
\TT{IJ}{KL}&=&
2\,\Gd\oversym{_{\phantom{2}}^{I[K}\,A_1^{L]J}\,}
+ \ft1{64}\,\GG^{IJKL}_{\dA\dB}\,A_3^{\dA\dB}
\;,
\non[1ex]
\TT{IJ}{A} &=& \GG^{[I}_{A\dA}\,A_2^{J]\dA}
\;,
\non[1ex]
\TT{A}{B}
&=&
\ft1{6144}\,\GG^{IJKL}_{AB}\,\GG^{IJKL}_{\dA\dB}\,A_3^{\dA\dB} 
+\Gth\,\Gd_{AB} \;.
\la{TinA}
\ea
These equations may be inverted and give the solution for the
tensors $A_{1,2,3}$ in terms of the $T$-tensor:
\ba
A_{1}^{IJ} &=& 
\ft87\,\Gth\,\Gd_{IJ}
+\ft1{7}\,\TT{IK}{JK}
\;,
\non[1ex]
A_{2}^{I\dA}&=&
-\ft17\,\GG^J_{A\dA}\,\TT{IJ}{A}
\;,
\non[1ex]
A_{3}^{\dA\dB}&=&
2\Gth\,\Gd_{\dA\dB}
+\ft1{48}\,\GG^{IJKL}_{\dA\dB}\,
\TT{IJ}{KL}
\;.
\la{AinT}
\ea

\subsection{Differential identities}

With the linear identities derived in the last section we may now
compute the variation of the tensors $A_{1,2,3}$ when $\cV$ is varied. 
Since the matrix $\cV$ lives in the adjoint representation, its
variation along an invariant vector field $\GS^A$ is given by
\be
\frac{\Gd\VV{\cM}{\cB}}{\Gd \GS^A} ~=~
f_{\cB}{}^{\cC A}\,\VV{\cM}{\cC} 
\quad \Longrightarrow \quad
\left\{ 
\begin{array}{rcl}
\displaystyle \frac{\Gd\VV{\cM}{IJ}}{\Gd \GS^A} &=& 
-\ft12\,\GG^{IJ}_{AB}\,\VV{\cM}{B} \\[2ex]
\displaystyle \frac{\Gd\VV{\cM}{B}}{\Gd \GS^A} &=& 
-\ft14\,\GG^{IJ}_{AB}\,\VV{\cM}{IJ}
\end{array}
\right. 
\;.
\la{varV}
\ee
{}From \Ref{AinT} we then obtain
\ba
\frac{\Gd A_1^{IJ}}{\Gd \GS^A} &=&
\ft1{14}\,
\left(\GG^{IK}_{AB}\,\TT{KJ}{B}+\GG^{JK}_{AB}\,\TT{KI}{B}\right)
\;,
\non
\frac{\Gd A_2^{I\dA}}{\Gd \GS^A} &=&
\ft1{14}\,\GG^J_{B\dA}\,
\left(\GG^{IJ}_{AC}\,\TT{B}{C} + \ft12\,\GG^{MN}_{AB}\,\TT{IJ}{MN}
\right)
\;,
\non
\frac{\Gd A_3^{\dA\dB}}{\Gd \GS^A} &=&
-\ft1{48}\,\GG^{IJKL}_{\dA\dB}\,\GG^{KL}_{\vphantom{\dA}AB}\;
\TT{IJ}{B}
\;.
\nn
\ea
Rewriting the expressions on the r.h.s. in terms of the tensors 
$A_{1,2,3}$ by means of \Ref{TinA} we get
\ba
\frac{\Gd A_1^{IJ}}{\Gd \GS^A} &=&
\GG^{(I}_{A\dA}\,A_2^{J)\dA} 
\;,
\non
\frac{\Gd A_2^{I\dA}}{\Gd \GS^A} &=&
\ft12\,\left(
\GG^M_{A\dA}\,A_1^{IM} + \GG^I_{A\dB}\,A_3^{\dA\dB} 
- \GG^I_{B\dA}\,\TT{A}{B} \right)
\;,
\non
\frac{\Gd A_3^{\dA\dB}}{\Gd \GS^A} &=&
\ft1{48}\,\GG^{IKMN}_{\dA\dB}\,\GG^{KMN}_{A\dC}\,
A_2^{I\dC}
\;.
\la{Avar}
\ea
This reproduces equations \Ref{varA} and \Ref{d2} from the last
chapter. In particular, we obtain the covariant derivatives of the
tensors $A_{1,2}$
\ba
\CD_\mu A_1^{IJ} &=&
\GG^{(I}_{A\dA}\,A_2^{J)\dA}\,\CP^A_\mu 
\;, 
\non
\CD_\mu A_2^{I\dA} &=&
\ft12\,\left(
\GG^M_{A\dA}\,A_1^{IM} + \GG^I_{A\dB}\,A_3^{\dA\dB} 
- \GG^I_{B\dA}\,\TT{A}{B} \right)\,\CP^A_\mu 
\;,
\la{Ader}
\ea
which coincide with equations \Ref{id2} found before. The variation
\Ref{Avar} further allows to compute the variation of the scalar
potential \Ref{potential} 
\ben
\frac{\Gd}{\Gd \Sigma^A} \Big( A_{1}^{IJ} A_{1}^{IJ}
-\ft12\,A_{2}^{I\dA} A_{2}^{I\dA} \Big) ~=~
\ft12\,\GG^M_{A\dA}\,\Big(
3A_1^{MN}\!A_2^{N\dA} - A_3^{\dA\dB}\!A_2^{M\dB}\Big) \;,
\een
which has also been used in the last chapter. Together with the
quadratic identity \Ref{Q4} to be derived below, this yields the
condition for stationary points of the potential
\be
\frac{\Gd W}{\Gd \GS^A}=0 
\qquad\Longleftrightarrow\qquad
3\,A_1^{IM}A_2^{M\dA} = A_3^{\dA\dB}A_2^{I\dB} \;.
\la{stationary}
\ee
Obviously, a sufficient condition for stationarity is $A_2^{I\dA}=0$\,.

\subsection{Quadratic identities}
So far, we have exploited the projector condition \Ref{crit} to derive
linear identities in $\TT{\cA}{\cB}$. However, additional information
stems from the fact that the tensor $\GTh_{\cM\cN}$ is built from
projectors onto subgroups, cf.~\Ref{emb}. This can be used to derive
further identities quadratic in the tensors $A_{1,2,3}$. As we have
seen in the previous chapter, identities of this type are also needed
to ensure supersymmetry of the gauged theory.

Since $\GTh_{\cM\cN}$ projects onto a subgroup $G_0\subset G$, it
satisfies:
\be\la{subgroup}
\GTh_{\cK(\cM}\,f_{\cN)}{}^{\cK\cL}\,\GTh_{\cL\cP} ~=~ 0 \;,
\ee
which follow from closure of $G_0$ and the antisymmetry of the 
structure constants. Invariance of the structure constants then implies
\be\la{Grel}
\GTh_{mn}\,\VV{n}{\cC}\;
f^{\cC\cD}{}^\vl_{(\cA}\,\TT{\cB)}{\cD}~=~0\;.
\ee
Evaluate this expression for $(\CA,\CB)=([IM],[KM])$:  
\ben
4\,\VV{m}{N(I}\,\TT{K)M}{MN}+  
\GG^{IM}_{AB}\,\VV{m}{A}\,\TT{KM}{B}+
\GG^{KM}_{AB}\,\VV{m}{A}\,\TT{IM}{B} ~=~0\;,
\een
where the index $m$ is projected onto the subalgebra
$\mfg_0$. Inserting \Ref{TinA} yields 
\be\la{Grel1}
\VV{m}{A} \, \GG^{(I}_{A\dA}\,A^{K)\dA}_2 = 
\VV{m}{IM}\,A_1^{MK} + \VV{m}{KM}\,A_1^{MI}\;,
\ee
and thus the identity \Ref{s1}, required above for closure of
the supersymmetry algebra in the gauged theory.  If we contract this
equation with $\VV{n}{JK}\GTh_{mn}$, symmetrize in $(IJ)$ and once
more insert \Ref{TinA}, we obtain
\be\la{Q1}
A_1^{IK}A_1^{KJ} - \ft12\,A_2^{I\dA}A_2^{J\dA}
= \ft1{16}\,\Gd^{IJ}\,
\Big(
A_1^{KL}A_1^{KL} - \ft12\,A_2^{K\dA}A_2^{K\dA}
\Big) \;.
\ee
This gives already the quadratic identity \Ref{qu1}. If on the other
hand we contract \Ref{Grel1} with $\GG^K_{B\dA}\VV{n}{B}\GTh_{mn}$, we
obtain after inserting \Ref{TinA}
\ba
\ft1{64}\,
\GG^{IKMN}_{\dC\dD}\,\GG^{MN}_{\dA\dB}\,A_2^{K\dB}A_3^{\dC\dD} \!
&=& \!
-32\,A_1^{IN}A_2^{N\dA}
+2\,(\GG^I\GG^K)_{\dA\dB}\,A_1^{KN}A_2^{N\dB} 
\non
&&{}+10\,A_2^{I\dB}A_3^{\dA\dB} 
-(\GG^I\GG^K)_{\dA\dB}\, A_2^{K\dC}A_3^{\dB\dC} 
\non
&&{}-16\,\Gth\,A_2^{I\dA}
\;.
\la{Q2}
\ea
Evaluating \Ref{Grel} for $(\CA,\CB)=([IJ],A)$ and contracting with
$\GG^J_{A\dA}$ leads to
\ba
\lefteqn{\ft1{6}\,(\GG^J\GG^{MNKL})_{A\dA}\,\VV{m}{A}\,\TT{MN}{KL}-
\ft1{12}\,(\GG^{MNKL}\GG^J)_{A\dA}\,\VV{m}{A}\,\TT{MN}{KL}  ~=} 
\non[.5ex]
&=&
\ft4{7}\,(\GG^K\GG^{MN})_{A\dA}\,\VV{m}{MN}\,\TT{JK}{A} 
-\ft{16}7\,\GG^K_{A\dA}\,\VV{m}{JM}\,\TT{MK}{A}
\non
&&{}+\ft87\,\GG^K_{A\dA}\,\VV{m}{A}\,\TT{JM}{MK} 
+\ft1{14}\,\GG^J_{A\dA}\,\VV{m}{A}\,\TT{MN}{MN} 
\;,
\la{Grel2}
\ea
again, if the index $m$ is projected onto the subalgebra $\mfg_0$. To
obtain the desired identity, we contract this equation with
$\VV{n}{IJ}\GTh_{mn}$ and insert \Ref{TinA}. After some calculation
we arrive at
\ba
\ft1{64}\,
\GG^{IKMN}_{\dC\dD}\,\GG^{MN}_{\dA\dB}\,A_2^{K\dB}A_3^{\dC\dD} 
&=&
64\,A_1^{IN}A_2^{N\dA}
-4\,(\GG^I\GG^K)_{\dA\dB}\,A_1^{KN}A_2^{N\dB} 
\non
&&{}-22\,A_2^{I\dB}A_3^{\dA\dB} 
+(\GG^I\GG^K)_{\dA\dB}\, A_2^{K\dC}A_3^{\dB\dC} 
\non
&&{}-16\,\Gth\,A_2^{I\dA}
\;.
\la{Q3}
\ea
Equating \Ref{Q2} and \Ref{Q3}, we finally obtain
\be\la{Q4}
3A_1^{IJ}A_2^{J\dA} - A_2^{I\dB}A_3^{\dA\dB} =
\ft1{16}\,(\GG^I\GG^J)_{\dA\dB}\,
\Big(3A_1^{JK}A_2^{K\dB} - A_2^{J\dC}A_3^{\dB\dC}\Big) \;.
\ee

We have thus shown that the condition \Ref{crit} together with the
fact that $\GTh_{\cA\cB}$ projects onto a subalgebra implies the
quadratic identities \Ref{Q1} and \Ref{Q4} which coincide with
\Ref{qu1}, \Ref{qu2} found above. Altogether, we recover in this fashion
all the identities required in Chapter~\ref{CHgauge} from the single 
projector condition \Ref{crit} for the embedding tensor $\GTh_{\cA\cB}$\,.

\mathon
\section{Admissible gauge groups $G_0$}
\mathoff
\la{CHG0}

Having reduced the consistency conditions required by local 
supersymmetry to a set of algebraic conditions \Ref{crit} 
for the embedding tensor of the gauge group $G_0\subset G$,
we must now ascertain  that this condition admits non-trivial 
solutions and classify them.  This is the objective of
the present section. As we will see the variety of solutions of
\Ref{crit}, each of which gives rise to a maximally supersymmetric
gauged supergravity, is far richer than in dimensions $D\geq 4$.

The power of equation \Ref{crit} is based on its formulation
as a single projector condition in the tensor product decomposition 
\Ref{sym248}. This permits the construction of solutions by purely 
group theoretical means. To demonstrate that these methods also
clarify the structure of the $T$-identities in $D\geq 4$, we derive 
the analog of \Ref{crit} to re-obtain the results of \cite{deWNic82} 
and \cite{GuRoWa86}. Group theoretical arguments then show immediately 
that the gauge groups $SO(8)$ and $SO(6)$, respectively, solve the
relevant equations. In particular, this provides a unifying argument 
for the consistency of all the noncompact gaugings found 
subsequently in \cite{Hull84a,Hull84b,GuRoWa86}.

The analysis for three dimensions turns out to be more involved, but
extending the above arguments we arrive at a variety of admissible gauge
groups. There is a regular series of gauge groups $SO(p,8\!-\!p)\cro
SO(p,8\!-\!p)$ including the maximal compact $SO(8)\cro SO(8)$, and 
several exceptional noncompact gauge groups, summarized in
Table~\ref{ncGH} below. Still this is not a complete classification 
of admissible gauge groups, as we restrict the analysis of compact
and noncompact gauge groups to the maximal subgroups of $SO(16)$ and
$\EE$, respectively. We leave the exploration of smaller rank
gauge groups to future work.

\mathon
\subsection{$T$-identities and gauge groups in higher dimensions}
\mathoff

As a ``warm-up'' let us first apply our techniques to the gauged
maximal supergravities in $D=4,5$. This will allow us to shortcut
the derivation of the (linear) $T$-identities given in the original
work.

\mathon
\subsubsection{$D\equ 4$}
\mathoff

Like \Ref{T}, the $D\equ 4$ $T$-tensor is obtained from a
constant $G_0$-invariant tensor $\GTh$ by a field dependent rotation
with the matrix $\cV\in E_{7(7)}$ in the fundamental
representation. The constant tensor $\GTh$ there transforms in the
product of the adjoint and the fundamental representation
\be
{\bf 56} \times {\bf 133}  ~=~  {\bf 56} + {\bf 912} + {\bf 6480}
\;,
\la{decD4}
\ee
of $E_{7(7)}$,\footnote{It is only for $\EE$ that the fundamental
representation coincides with the adjoint representation and the
tensor $\GTh$ hence coincides with the embedding tensor of the group
$G_0$.}  such that $T$ is cubic rather than quadratic in the matrix
entries of $\cV$. Computations similar to those presented in the last
chapter then show that full supersymmetry of the gauged Lagrangian is
equivalent to
\be
T= T^{\bf 912} \quad\Longleftrightarrow\quad \GTh=\GTh^{\bf 912} \;, 
\la{critD4}
\ee
providing the analogue of \Ref{crit}. It is now straightforward to see
that $G_0=SO(8)$ indeed gives a solution to \Ref{critD4}: consider
the decomposition of \Ref{decD4} under $SO(8)$:\,\footnote{LiE
\cite{LeCoLi92} has been very helpful to quickly determine these
decompositions.} 
\ba
{\bf 56} &\ra & 2\cdot 28 \;,\non
{\bf 912} &\ra & 2\cdot 1 + 2\cdot 35_v +  
2\cdot 35_s +  2\cdot 35_c + \dots \;, \non
{\bf 6480} &\ra& 6\cdot 28 + 2\cdot 35_v +  
2\cdot 35_s +  2\cdot 35_c + \dots \;.
\ea
As the singlets appear only in the ${\bf 912}$, any
$SO(8)$ invariant tensor in \Ref{decD4} automatically satisfies
\Ref{critD4}. The same argument proves the consistency of the
noncompact $SO(p,8\!-\!p)$ gaugings found in \cite{Hull84b}. As
shown in \cite{CFGTT98} equation \Ref{critD4} indeed contains 
no other solutions than those found in \cite{deWNic82,Hull84b}.

\mathon
\subsubsection{$D\equ 5$}
\mathoff
For $D=5$, the constant tensor $\GTh$ transforms in the product 
of the adjoint and the fundamental representation
\be
{\bf 27} \times {\bf 78}  ~=~  {\bf 27} + {\bf 351} + {\bf 1728}
\;,
\la{decD5}
\ee
%
of $E_{6(6)}$.
Rotation by $\cV$ in the fundamental
representation of $E_{6(6)}$ converts $\GTh$ into the $T$-tensor, cubic
in the matrix entries of $\cV$. Supersymmetry of the gauged Lagrangian
then is shown to be equivalent to  
\be
T= T^{\bf 351} \quad\Longleftrightarrow\quad \GTh=\GTh^{\bf 351} 
\;, 
\la{critD5}
\ee
in analogy with \Ref{crit} and \Ref{critD4}. Again, it is 
straightforward to see that $G_0=SO(6)$ yields a solution to
\Ref{critD5}: under $SO(6)$, \Ref{decD5} decomposes as
%
\ba
{\bf 27} &\ra & 2\cdot 6 + 15 \;,\non
{\bf 351} &\ra & 1 +  2\cdot6 + 2\cdot10 + 2\cdot\overline{10} 
+ 4\cdot 15 + \dots \;, \non
{\bf 1728} &\ra& 10\cdot 6 +  2\cdot10 + 2\cdot\overline{10} 
+ 9\cdot 15 +\dots \;.
\ea
Now the singlet appears only in the ${\bf 351}$, hence there is just one
$SO(6)$ invariant tensor in \Ref{decD5} which automatically satisfies
\Ref{critD5}. As before, this argument generalizes to all the noncompact
gauge groups found in \cite{GuRoWa86}.

\subsection{Compact gauge groups}

Let us now come back to \Ref{crit}. We will first consider compact
gauge groups $G_0\subset SO(16)$. Their embedding tensors satisfy
\be\la{cp}
\GTh_{IJ,A} = 0 = \GTh_{A,B} \;;
\ee
the only nonvanishing component is $\GTh_{IJ,KL}$ which under $SO(16)$
decomposes as
\be
\GTh_{IJ,KL} ~\sim~ 1 + 135 + 1820 + 5304 \;.
\ee
According to \Ref{break}, the $5304$ is part of the ${\bf 27000}$ and
must vanish for \Ref{crit} to be satisfied. From \Ref{critex} it
further follows that the 1 and the 1820 coincide with the
corresponding parts in $\GTh_{A,B}$ and thus must vanish due to
\Ref{cp}. Hence, for compact $G_0$, only the $135$ representation
survives, and the condition \Ref{crit} reduces to
\be
\GTh_{IJ,KL} = \Gd\undersym{_{I[K}\,\Xi_{L]J}\,} \;,
\qquad \mbox{with}\quad
\Xi_{IJ} = \ft72\,\GTh_{IK,JK}\;,\quad \Xi_{II}=0  \;.
\la{critcp}
\ee
The tracelessness of $\GTh$ in particular rules out any simple compact
gauge group. 

In principle, the elementary form of the constraint \Ref{critcp}
should allow a complete classification of the possible compact gauge
groups; however, in the following, we restrict attention to the
maximal subgroups of $SO(16)$. They are 
\ba
SO(9)\;,\quad SO(5)\cro SO(5)\;, \quad
SO(3)\cro U\!Sp(8)\;, \non
\mbox{and} \quad
SO(p)\cro SO(16\!-\!p)\;,\quad \mbox{for}\quad p=0, \dots, 8 \;. 
\la{maxcp}
\ea
A necessary condition for a compact gauge group to be admissible
immediately follows from \Ref{critcp}: there must exist a
$G_0$-invariant tensor $\Xi_{IJ}$ in the 135 of $SO(16)$. In other
words, there must be a singlet in the decomposition of 135 w.r.t.\
$G_0$. From the maximal subgroups \Ref{maxcp} this already rules out
the first three.  It remains to study the
$SO(p)\!\times\!SO(16\!-\!p)$. These groups have a unique invariant
tensor in the 135:
\be
\Xi_{ij} = (16\!-\!p)\,\Gd_{ij}\;,\quad 
\Xi_{\ovi \ovj} = -p\,\Gd_{\ovi \ovj}  \;,
\ee
where $i,j =1, \dots p$ and $\ovi,\ovj =p\!+\!1, \dots, 16$ denote the
splitting of the $SO(16)$ vector indices $I$, and the relative factor
between $\Xi_{ij}$ and $\Xi_{\ovi \ovj}$ is determined from
tracelessness. By \Ref{critcp}, the tensor $\GTh_{IJ,KL}$
satisfying \Ref{crit} is
\ben
\GTh_{ij,kl} = (16\!-\!p)\,\Gd^{ij}_{kl}
\;,\quad
\GTh_{\ovi\ovj,\ovk\ovl} = 
-p\,\Gd^{\ovi\ovj}_{\ovk\ovl}
\;,\quad
\GTh_{i\ovj,k\ovl} = \ft12(8\!-\!p)\,\Gd_{ik}\,\Gd_{\overline{jl}}
\;.
\een
However, due to the nonvanishing mixed components
$\GTh_{i\ovj,k\ovl}$, this tensor coincides with the embedding tensor
of $SO(p)\!\times\!SO(16\!-\!p)$ if and only if $p\equ 8$. Hence we
have shown that the only maximal subgroup of $SO(16)$ whose embedding
tensor satisfies the condition \Ref{crit} is
\be\la{88}
G_0 = SO(8)\cro SO(8) \subset SO(16) \;,
\ee
where the ratio of coupling constants of the two factors is
$\cc{1}/\cc{2}=-1$; in particular the trace part $\Gth$ of
$\GTh_{\cA\cB}$ vanishes. Combining this with the results of the 
previous chapters, we have thus shown the existence of a maximally
supersymmetric gauged supergravity with compact gauge group
$G_0=SO(8)\cro SO(8)$. Under $G_0$, the scalar degrees of freedom
decompose as
\be
120\ra (1,28)\!+\!(28,1)\!+\!(8_s,8_c)  \;,\quad
128\ra (8_v,8_v)\!+\!(8_c,8_s) \;,
\la{88bos}
\ee
while the spinors split into
\be
16\ra (1,8_c)\!+\!(8_s,1)  \;,\qquad
\overline{128}\ra (8_v,8_s)\!+\!(8_c,8_v) \;.
\la{88fer}
\ee
Amongst other things we here recognize the standard decomposition of 
the on-shell IIA supergravity multiplets in terms of left and right 
moving string states.

\subsection{Regular noncompact gauge groups}

In order to identify the allowed noncompact gauge groups, we first
recall that for the maximal gauged supergravity in $D\equ 4$, 
several noncompact gaugings were found by analytic continuation
\cite{Hull84a,Hull84b}. The noncompact gauge groups are thus
alternative real form of the complexified gauge group 
$SO(8,{\mathbb{C}})$, and the consistency of the noncompact
gaugings was basically a consequence of the consistency of
the original theory \cite{deWNic82} with compact gauge group. 
The results of the last section suggest that analogous gaugings 
should exist for the different real forms of \Ref{88}.

The complexification of \Ref{88} is $SO(8,{\mathbb{C}})\cro
SO(8,{\mathbb{C}})$. Its real forms which are also contained 
in $\EE$ are given by
\be
G_0=SO(p,8\!-\!p)^{(1)}\cro SO(p,8\!-\!p)^{(2)} \;,\quad
\mbox{for}\;\; p=1, \dots, 4 \;.
\la{ncreg}
\ee
They are embedded in $\EE$ via the maximal noncompact subgroup
$SO(8,8)$. Therefore the latter group is the analogue of the
subgroups $SL(8,{\mathbb{R}})\subset E_{7(7)}$ in $D\!=\!4$ and
$SL(6,{\mathbb{R}})\cro SL(2,{\mathbb{R}})\subset E_{6(6)}$ in
$D\!=\!5$. To further illustrate the embedding, we have denoted the
two factors of $G_0$ by superscripts $\scriptstyle{(1)}$,
$\scriptstyle{(2)}$ whereas we denote the two factors of \Ref{88} by
subscripts $\scriptstyle{L,R}$. The maximal compact subgroup of
\Ref{ncreg} is given by
\ba
H_0&=&H^{(1)}\times H^{(2)} \non[1ex]
&\equiv&
\Big(SO(p)^{(1)}_L\cro SO(8\!-\!p)^{(1)}_R \Big) \times
\Big(SO(p)^{(2)}_R\cro SO(8\!-\!p)^{(2)}_L \Big) 
\;,
\la{Hpq}
\ea
with
\ba
H^{(1)} \subset SO(p,8\!-\!p)^{(1)}\;,&&
SO(p)^{(1)}_L\cro SO(8\!-\!p)_L^{(2)} \subset SO(8)_L \;,
\non
H^{(2)} \subset SO(p,8\!-\!p)^{(2)}\;,&&
SO(p)^{(1)}_R\cro SO(8\!-\!p)_R^{(2)} \subset SO(8)_R \;.
\nn
\ea
The embedding of $H_0$ into $SO(8)_L\cro SO(8)_R$ is the standard one,
without any triality rotation. In other words, the $8_v$ of $SO(8)_L$ 
decomposes into $(p,1)+(1,8\!-\!p)$ under $SO(p)^{(1)}_L\cro
SO(8\!-\!p)_L^{(2)}$, etc.

Consistency of the gauged theories with noncompact gauge groups
\Ref{ncreg} could in principle be shown in analogy with
\cite{Hull84b,HulWar85} by the method of analytic continuation. 
Alternatively, their consistency follows from an algebraic
argument along the lines of the last section by use of our form of
the consistency condition \Ref{crit}. This gives the analogue of the
noncompact gaugings found in higher dimensions
\cite{Hull84b,GuRoWa86}.

\subsection{Exceptional noncompact gauge groups}

Next, we discuss noncompact gauge groups, which unlike the groups 
identified in \Ref{ncreg} do not share the complexification 
with any compact subgroup contained in $\EE$. Their existence 
is again a consequence of the absence of any {\it a priori} 
restriction on the number of vector fields in three dimesnions.

These noncompact solutions to \Ref{crit} may be found by a purely
group theoretical argument. As an example, consider the maximal 
subgroup $G_0 = G_{2(2)}\!\times\!F_{4(4)}$. Under $G_0$ the adjoint
representation of $\EE$ decomposes as
\be
{\bf 248} ~\ra~ (14,1) + (1,52) + (7,26) \;.
\ee
Accordingly, the symmetric tensor product \Ref{sym248} contains three
singlets under $G_0$, and the Cartan-Killing form of $\EE$
decomposes into three $G_0$-invariant tensors:
\be
\la{etaE}
\eta_{\cM\cN} ~=~ 
\eta^{(14,1)}_{\cM\cN}+\eta^{(1,52)}_{\cM\cN}+\eta^{(7,26)}_{\cM\cN} 
\;.
\ee
More precisely, each of the three terms on the r.h.s.\ of
\Ref{sym248} contains exactly one singlet under
$G_0$~\cite{LeCoLi92}. Consequently,  there is a linear combination
\ben
\Ga_1\, \eta^{(14,1)}_{\cM\cN} + \Ga_2\, \eta^{(1,52)}_{\cM\cN} 
+ \Ga_3\, \eta^{(7,26)}_{\cM\cN} \;,
\een
which lies entirely in the ${\bf 3875}$. Subtracting a proper multiple
of the $\EE$ singlet \Ref{etaE}, we find that
\be
\GTh_{\cM\cN} ~\equiv~
(\Ga_1\!-\!\Ga_3)\,\eta^{(14,1)}_{\cM\cN} + 
(\Ga_2\!-\!\Ga_3)\,\eta^{(1,52)}_{\cM\cN} \;,
\la{THFG}
\ee
satisfies \Ref{crit}. This is the embedding tensor of $G_0 =
G_{2(2)}\!\times\!F_{4(4)}$ with a fixed ratio of coupling constants
between the two factors, which solves \Ref{crit} and \Ref{subgroup}. 
The results of the last chapter then prove the existence of a maximally 
supersymmetric gauged theory with gauge group $G_{2(2)}\!\times\!F_{4(4)}$. 

The same argument may be applied to other noncompact subgroups of
$\EE$. A closer inspection of the above proof reveals that only two
ingredients were needed, namely ($i$) that the gauge group $G_0$
consists of two simple factors and ($ii$) that the $\EE$
representations ${\bf 3875}$ and ${\bf 27000}$ each contain precisely
one singlet in the decomposition under $G_0$. As it turns out, this
requirement is also met by the noncompact groups
$E_{7(7)}\!\times\!SL(2)$, $E_{6(6)}\!\times\!SL(3)$, and all their
real forms which are contained in $\EE$. The list of exceptional
noncompact subgroups passing this test, together with their maximal
compact subgroups is displayed in Table \ref{ncGH}.

\begin{table}[htb]
\centering
\vspace{2ex}
\begin{tabular}{||c|c||} 
\hline
&\\[-1.8ex]
$G_0=G^{(1)}\cro G^{(2)}$
&
maximal compact subgroup $H_0=H^{(1)}\cro H^{(2)}$
\\[.5ex]
\hline
\hline
&\\[-1.8ex]
$G_{2(2)}\!\times\!F_{4(4)}$ 
&
$\Big(SU\!(2)_L\cro SU\!(2)_R\Big)\cro 
\Big(SU\!(2)_L\cro U\!Sp(6)_R\Big)$ \\ [1.2ex]
\hline
&\\[-1.8ex]
$G_{2}\times F_{4(-20)}$ 
&
$\vphantom{\Big(}(G_{2})_L\cro SO(9)_R$ \\[1.2ex]
\hline
&\\[-1.8ex]
$E_{6(6)}\!\times\!SL(3)$
&
$\vphantom{\Big(}U\!Sp(8)_L\times SU\!(2)_L$ \\[1.2ex] 
\hline
&\\[-1.8ex]
$E_{6(2)}\!\times\!SU(2,1)$
&
$\Big(SU\!(6)_L\cro SU\!(2)_R\Big)\cro 
\Big(SU\!(2)_R\cro U\!(1)_L\Big)$ \\[1.2ex] 
\hline
&\\[-1.8ex]
$E_{6(-14)}\!\times\!SU(3)$
&
$\Big(SO(10)_L\cro U\!(1)_R\Big)\cro SU\!(3)_R$ \\[1.2ex] 
\hline
&\\[-1.8ex]
$E_{7(7)}\!\times\!SL(2)$ 
&
$\vphantom{\Big(}SU\!(8)_L\times U\!(1)_L$ \\ [1.2ex]
\hline
&\\[-1.8ex]
$E_{7(-5)}\!\times\!SU(2) $ 
&
$\Big(SO(12)_L\cro SU\!(2)_R\Big)\cro SU\!(2)_R$ \\[1.2ex]
\hline
&\\[-1.8ex]
$E_{8(8)}$
&
$\vphantom{\Big(}SO(16)_L$
\\[.8ex]
\hline
\end{tabular}
\bf\small
\caption{{\rm\small Exceptional noncompact gauge groups and their maximal
compact subgroups. The subscripts $\scriptstyle{L}$ and
$\scriptstyle{R}\;$ refer to the AdS supergroups $G_L\cro G_R$
associated to the maximally supersymmetric groundstates of these
theories, see Chapter~\ref{CHgs}.}}
\label{ncGH}      
\end{table}

There are also real forms of these exceptional gauge groups --- the
compact forms of $E_d$ for $d\equ 6, 7, 8$, and the real forms
$E_{8(-24)}, E_{7(-25)}$ and $E_{6(-26)}$ --- which are not contained
in $\EE$ and thus do not appear in this list. However, every real form
that may be embedded in $\EE$ gives rise to a maximally supersymmetric
gauged supergravity. The ``extremal'' noncompact solution to
\Ref{crit} is given by the group $G_0=\EE$ itself, in which case
$\GTh_{\cA\cB}$ reduces to the Cartan-Killing form~$\eta_{\cA\cB}$.

To complete the construction of the theories with gauge groups given
in Table~\ref{ncGH}, it remains to compute the ratio of coupling
constants between the two factors of $G_0$ which came out to be fixed
to a specific value in \Ref{THFG}. To this end, let us consider the
general situation of a gauge group with two simple factors
$G_0=G^{(1)}\cro G^{(2)}$, such that its maximal compact subgroup
likewise factors as $H_0=H^{(1)}\cro H^{(2)}$. Denote the embedding
tensor of $G_0$ by
\be
g \GTh_{\cM\cN} ~=~ g_1\,\eta^{(1)}_{\cM\cN} + g_2\,\eta^{(2)}_{\cM\cN}
\;, 
\la{THGG}
\ee
where $\eta^{(1),(2)}$ are the embedding tensors of $G^{(1),(2)}$,
respectively, and assume that \Ref{THGG} satisfies
\Ref{crit}. Equation \Ref{THFG} was a particular case satisfying these
assumptions. Contracting \Ref{THGG} with $\eta^{\cM\cN}$ yields
\ben
g \Gth\;{\rm dim~} \EE ~=~ 
g_1\,{\rm dim~} G^{(1)} + g_2\,{\rm dim~} G^{(2)}
\;. 
\een
where the l.h.s.\ follows from \Ref{TH123}.  On the other hand,
contracting \Ref{THGG} with $\eta^{IJ,KL}$ over the compact part of
$\EE$ gives
\ben
g \Gth\;{\rm dim~} SO(16) ~=~ 
g_1\,{\rm dim~} H^{(1)} + g_2\,{\rm dim~} H^{(2)}
\;,
\een
where the l.h.s.\ here follows from \Ref{critex} --- and is a
consequence of the fact that due to \Ref{crit} the only $SO(16)$
singlet in $\GTh_{\cM\cN}$ is given by the first term in \Ref{TH123}.

{}From the last two equations one may extract the coupling constants
$g_1$, $g_2$ of the two factors of the gauge group. Their ratio is
\be
\frac{g_1}{g_2} ~=~ 
-\frac{15\, {\rm dim~} G^{(2)}-31\, {\rm dim~} H^{(2)}}
{15\, {\rm dim~} G^{(1)}-31\, {\rm dim~} H^{(1)}}
\;\;.
\la{ratio}
\ee
With the gauge groups and their compact subgroups given in
Table~\ref{ncGH} we then immediately obtain the ratios of coupling
constants for all these groups. In particular, no degeneration occurs
where this ratio would vanish or diverge.  In Table \ref{ncgroups}
displayed in the introduction, we have presented a list of all the
noncompact admissible subgroups $G_0\subset\EE$ , together with their
ratio of coupling constants. Remarkably, the ratios as determined by
\Ref{ratio} come out to be independent of the particular real form for
each of these exceptional noncompact groups. This suggests that the
theories whose gauge groups are different real forms of the same
complexified group may be related by analytic continuation, in a 
similar fashion as the $SO(p,8\!-\!p)$ gaugings of the $D=4$ theory 
are related via $SO(8,{\mathbb C})$ \cite{Hull84a,Hull84b,HulWar85}. 
Here, the analytic continuation would have to pass through the 
complex group $E_8({\mathbb C})$.

This concludes our discussion of admissible gauge groups. We note that
in addition to the groups identified in this chapter there should also
exist non-semisimple gaugings analogous to the theories constructed
in \cite{Hull84a,Hull84b,HulWar85,ACFG01}. We leave their exploration
and complete classification for future study.

\section{Stationary points with maximal supersymmetry}
\la{CHgs}

The point of vanishing scalar fields, i.e. $\cV\equ I$, plays a
distinguished role: it is a stationary point with maximal
supersymmetry for all the theories we have constructed. Recall that
the condition for stationarity was already spelled out in \Ref{stationary}.
At $\cV\equ I$, the gauge group $G_0$ is broken to its maximal compact
subgroup $H_0$. For the compact gauge group \Ref{88}, the tensor
$A_2^{I\dA}$ vanishes at this point, since $\GTh$ has no contribution
in the noncompact directions, cf.~\Ref{AinT} and \Ref{cp}. Hence,
\Ref{stationary} is satisfied; the compact gauged theory has a $G_0$
invariant stationary point at $\cV\equ I$.  For the noncompact real
forms \Ref{ncreg}, the decomposition \Ref{88fer} implies that there is
no $H_0$-invariant tensor in the tensor product
$16\cro\overline{128}$; hence, $A_2^{I\dA}$ vanishes also in these
theories at $\cV\equ I$. The same argument works also for the 
exceptional noncompact gauge groups from Table~\ref{ncgroups}. 
In summary, all the three-dimensional theories we have 
constructed share the stationary point $\cV\equ I$.

If we denote by $\nu\equ{\rm dim~} G_0$ and $\Gk\equ{\rm dim~} H_0$
the dimension of the gauge group and its maximal compact subgroup,
respectively, the field equations \Ref{NAduality} imply that for
$\cV\equ I$ the vector fields split into $\nu\!-\!\Gk$ massive
self-dual vectors and a $H_0$-Chern-Simons theory of $\Gk$ vector fields
which do not carry propagating degrees of freedom. In this way, the
erstwhile topological vector fields corresponding to the noncompact
directions in $G_0$ acquire a mass term by a Brout-Englert-Higgs like
effect as observed in \cite{DesYan89}.  Dropping the massive vector
fields as well as the matter fermions, the theory then reduces to a
$H_0$ CS theory, coupled to supergravity. Since the AdS$_3$
(super-)gravity itself allows the formulation as a CS theory of the
AdS group $SO(2,2)$ \cite{AchTow86,Witt88}, the resulting theory is a
CS theory with connection on a superextension of $H_0\cro SO(2,2)$. We
shall determine these supergroups in the following.

In order to analyze the residual supersymmetries at the stationary 
point $\cV\equ I$ in a little more detail, we consider the Killing
spinor equations, derived from \Ref{susyf}, \Ref{ferm} in absence of
the vector fields:
\ba
0&\stackrel{!}{=}&
\dd^{\vphantom b}_\mu \Ge^I + 
\ft12\,\I \Gg_a \left( A_\mu{}^a \,\Gd^{IJ} - 2g \,e_\mu{}^a
\,A_1^{IJ}\right) \,\Ge^J 
\;,
\la{KS1}\\
0&\stackrel{!}{=}& 
A_2^{I\dA} \,\Ge^I
\;.
\la{KS2}
\ea
%
Adapting the arguments of \cite{GuRoWa86} to the present case, it may
be shown that \Ref{KS1} in fact implies \Ref{KS2}. Namely, comparing
\Ref{KS1} to \Ref{covAdS} we find that every solution to \Ref{KS1}
corresponds to the product of an AdS$_3$ Killing spinor and an
eigenvector $\Ge_0^I$ of the real symmetric matrix $A_1^{IJ}$; the
eigenvalue $\alpha_i$ of $A_1^{IJ}$ is related to the AdS radius by
\be
2g\,|\alpha_i|= m \;.
\la{unbroken}
\ee
%
On the other hand, the Einstein field equations derived from
\Ref{L0123} imply that
\be
R_{\mu\nu} = 4 W_0 \, g_{\mu\nu} \;,
\ee
where $W_0$ is the value of the potential \Ref{potential} at the
critical point. From \Ref{Einstein} we infer the relation
$m^2=2W_0$. Given the eigenvector $\Ge_0^I$ of $A_1^{IJ}$ with
eigenvalue $\alpha_i$, we contract \Ref{qu1} with $\Ge_0^I$ to obtain
\be
\left(2\,g^2\,\alpha_i^2 - W_0 \right)\Ge_0^I  ~=~ 
g^2\,A_2^{I\dA}A_2^{J\dA} \Ge_0^J \;.
\la{stsusy}
\ee
If $\alpha_i$ satisfies \Ref{unbroken}, this equation indeed implies
\Ref{KS2}. As in higher dimensions, the number of residual
supersymmetries therefore corresponds to the number of eigenvalues 
$\alpha_i$ of $A_1^{IJ}$ satisfying \Ref{unbroken}.
Conversely, equation \Ref{stsusy} shows that $A_2^{I\dA}=0$ is a
sufficient condition for a maximally supersymmetric ground state: 
all eigenvalues of the tensor $A_1^{IJ}$ then satisfy \Ref{unbroken},
splitting into $16=n_L\!+\!n_R$ with positive and negative sign,
respectively. Altogether, we have thus shown that all the
theories with noncompact gauge groups from \Ref{ncreg} and
Table~\ref{ncgroups} possess a maximally supersymmetric ground state
at $\cV\equ I$. This is in marked contrast to the higher-dimensional
models, where several of the noncompact gaugings do not even admit
any stationary points \cite{Hull84a,Hull84b,HulWar85}.

Not unexpectedly, the background isometries of these groundstates are 
superextensions of the three-dimensional AdS group $SO(2,2)$. Since
$SO(2,2)=SU(1,1)_L\cro SU(1,1)_R$ is not simple, they are
in general direct products of two simple supergroups $G_L\cro
G_R$. Accordingly, the sixteen supersymmetry generators split into
$N=(n_L,n_R)$, such that the groups $G_{L,R}$ are $n_{L,R}$
superextensions of the $SU(1,1)_{L,R}$ with bosonic subgroups
\be
G_{L,R} ~\supset~ H_{L,R} \cro SU(1,1)_{L,R} \;.
\la{adsbos}
\ee
A list of possible factors $G_{L,R}$ based on the classification
\cite{Kac77,Nahm78} is given in \cite{GuSiTo86}.

To determine the AdS supergroups $G_L\cro G_R$ corresponding to the
maximally supersymmetric ground state of the theory with gauge group
$G_0$, one must identify the groups $H_{L,R}$ among the simple factors
of its maximally compact subgroup $H_0$, such that $H_0 = H_L\cro
H_R$. This basically follows from the decomposition of the sixteen
supercharges under $H_0$.  Note that $H_L$ is not necessarily entirely
contained in one of the two factors of the semisimple gauge group
$G_0$. Rather we find that in the two factorizations
\be
H^{(1)}\cro H^{(2)} ~=~ H_0 ~=~ H_L\cro H_R \;,
\la{H12LR}
\ee
the various subfactors are distributed in different ways among the two
factors. This has been made explicit in \Ref{Hpq} and
Table~\ref{ncGH}, respectively, by designating the simple factors of
$H_0$ with the corresponding sub- and superscripts. In fact, the only
gauge groups for which the two factorizations \Ref{H12LR} coincide are
the compact group \Ref{88}, the group $G_{2}\cro F_{4(-20)}$ from
Table~\ref{ncGH}, and the gauge group $\EE$ itself. For the noncompact
gauge groups $E_{6(6)}\!\times\!SL(3)$, $E_{7(7)}\!\times\!SL(2)$, and
$E_{8(8)}$, we find $H_0\equ H_L$, i.e.\ $G_R$ reduces to its purely
bosonic AdS part $SU(1,1)_R$. Another particular situation arises for
the noncompact gauge group $SO(4,4)\cro SO(4,4)$, where the
supergroups $G_{L,R}$ themselves are not simple but direct products of
two supergroups, respectively.

The complete list is given in Table~\ref{background}, where we have
summarized the background isometries of the maximally supersymmetric
stationary point $\cV\equ I$ for all the three-dimensional gauged
maximal supergravities constructed in this article.

\begin{table}[htb]
\centering
\vspace{2ex}
\begin{tabular}{||l|c|l||} 
\hline
&&\\[-1.8ex]
gauge group $G_0$
&
$N=(n_L,n_R)$
&
background supergroup $G_L \cro G_R$\\[.5ex]
\hline
\hline
&&\\[-1.8ex]
$SO(8)\cro SO(8)$
&
$(8,8)$ &
$O\!Sp(8|2,{\mathbb R})\cro O\!Sp(8|2,{\mathbb R})$\\[.5ex]
\hline
&&\\[-1.8ex]
$SO(7,1)\cro SO(7,1)$
&
$(8,8)$ &
$F(4)\cro F(4)$\\[.5ex]
\hline
&&\\[-1.8ex]
$SO(6,2)\cro SO(6,2)$
&
$(8,8)$ &
$SU(4|1,1)\cro SU(4|1,1)$\\[.5ex]
\hline
&&\\[-1.8ex]
$SO(5,3)\cro SO(5,3)$
&
$(8,8)$ &
$O\!Sp(4^*|4)\cro O\!Sp(4^*|4)$\\[.5ex]
\hline
&&\\[-1.8ex]
$SO(4,4)\cro SO(4,4)$
&
$(8,8)$ &
$\left[\;D^1(2,1;-1) \cro SU(2|2)\;\right]^2$\\[.5ex]
\hline
&&\\[-1.8ex]
$G_{2(2)}\!\times\!F_{4(4)}$ 
&
$(4,12)$ &
$D^1(2,1;-\ft23) \cro O\!Sp(4^*|6)$ \\[.5ex]
\hline
&&\\[-1.8ex]
$G_{2}\!\times\!F_{4(-20)}$ 
&
$(7,9)$ &
$G(3)\cro O\!Sp(9|2,{\mathbb R})$ \\[.5ex]
\hline
&&\\[-1.8ex]
$E_{6(6)}\!\times\!SL(3)$
&
$(16,0)$&
$O\!Sp(4^*|8)\cro SU(1,1)$ 
\\[.5ex] 
\hline
&&\\[-1.8ex]
$E_{6(2)}\!\times\!SU(2,1)$
&
$(12,4)$
&
$SU(6|1,1) \cro D^1(2,1;-\ft12) $
\\[.5ex] \hline
&&\\[-1.8ex]
$E_{6(-14)}\!\times\!SU(3)$
&
$(10,6)$
&
$O\!Sp(10|2,{\mathbb R})\cro SU(3|1,1)$
\\[.5ex] 
\hline
&&\\[-1.8ex]
$E_{7(7)}\!\times\!SL(2)$ 
&
$(16,0)$ &
$SU(8|1,1)\cro SU(1,1)$ 
\\[.5ex]
\hline
&&\\[-1.8ex]
$E_{7(-5)}\!\times\!SU(2) $ 
&
$(12,4)$ &
$O\!Sp(12|2,{\mathbb R}) \cro D^1(2,1;-\ft13)$
\\[.5ex]
\hline
&&\\[-1.8ex]
$E_{8(8)}$
&
$(16,0)$
&
$O\!Sp(16|2,{\mathbb R})\cro SU(1,1)$ \\[.5ex]
\hline
\end{tabular}
\bf\small
\caption{{\rm\small Background isometries of the maximally
supersymmetric ground states}}
\label{background}      
\end{table}

Let us emphasize that this table presumably represents only the 
tip of the iceberg as we expect there to be a wealth of 
stationary points with partially broken supersymmetry for 
``small'' gauge groups $G_0\subset E_{8(8)}$. On the other hand, 
for ``large'' gauge groups stationary points will be more scarce. 
As a special example, consider the extremal theory with
noncompact gauge group $\EE$, for which the potential becomes just 
a (cosmological) constant, and does not exhibit any stationary 
points besides the trivial one. In this case $\cV\equ I$ may always
be achieved by gauge fixing the local $\EE$ symmetry. Even after
this gauge fixing, by which the scalar fields have been eliminated 
altogether, there still remains the ``composite'' local $SO(16)$ 
invariance rendering 120 vectors out of the 248 vector fields 
unphysical. Accordingly, the theory in this gauge may be 
interpreted as an $SO(16)$ Chern-Simons theory coupled to
128 massive selfdual vector fields, each of which represents one
physical degree of freedom. In other words, with respect to the ungauged
theory, the propagating degrees of freedom have been shifted from 
the scalar fields to massive selfdual vectors. This is in fact an
extremal case of the mechanism required for gauging higher dimensional
supergravities in odd dimensions
\cite{ToPivN84,PePivN84,PePivN85,GuRoWa86} whereby massless $k\!-\!1$
forms in a $2k\!+\!1$ dimensional space-time upon gauging turn into
massive selfdual $k$-forms.  As discussed above, truncating the
massive vector fields together with the matter fermions, the theory
reduces to the $O\!Sp(16|2,{\mathbb R})$ theory of \cite{AchTow86} and
reproduces its $(16,0)$ supersymmetric ground state.

It will be most interesting to study the boundary theories associated 
with the gauged supergravities. The background isometries given 
in Table~\ref{background} determine the superconformal symmetries 
of the theories on the AdS$_3$ boundary. The chiral algebras
are obtained by Hamiltonian reduction of the current algebras based on the
AdS$_3$ supergroups $G_L$ and $G_R$, respectively (see \cite{dBoe99} for a
discussion and a translation table). For instance, the boundary theory 
of the superextended Chern-Simons theories \cite{AchTow86} is described
by a super-Liouville action with $SO(n)$ extended superconformal
symmetry \cite{CoHevD95,HeMaSc00}. The maximal gauged supergravities
\Ref{L0123} then introduce additional scalar and massive vector degrees 
of freedom, respectively, which propagate in the bulk.

\section{Outlook: a higher dimensional ancestor?}

As already pointed out in the introduction there appears to be no way
to obtain the gauged models constructed in this paper by means of a
conventional Kaluza Klein compactification, because the latter would
give rise to a standard Yang-Mills-type Lagrangian with a kinetic term
for the vector fields, instead of the CS term that was required
here. Moreover, $D\!=\!11$ supergravity does not admit maximally
supersymmetric groundstates of the type AdS$_3 \times {\cal M}_8$ (see
e.g. \cite{DuNiPo86}), and even if it did, there simply are no
8-manifolds ${\cal M}_8$ whose isometry groups would coincide with the
gauge groups $G_0$ that we have found (since there are no 7-manifolds
with these isometries either, the arguments {\it a fortiori} also
excludes type-IIB theory as a possible ancestor).  Nonetheless all
these gauged models constitute continuous deformations of the original
$N\!=\!16$ theory of \cite{MarSch83}, which itself {\it is} derivable
by a torus reduction of $D\!=\!11$ supergravity.  The situation is
therefore quite different from the one in dimensions $D\!\geq \! 4$
where the gauged theories do emerge via sphere compactifications of
$D\!=\!11$ supergravity.\footnote{For the AdS$_4 \times S^7$
compactification this was rigorously shown in \cite{deWNic87}, while
for the AdS$_7 \times S^4$ a complete proof was given more recently
\cite{NaVavN99}. By contrast, the full consistency of the AdS$_5
\times S^5$ truncation of IIB supergravity remains an open problem
despite much supporting evidence, see \cite{PilWar00} and references
therein.}  This raises the question whether there exists a
higher-dimensional ancestor theory that would give rise to these
theories, and if so, what it might be. While we have no answer to this
question at the moment, we would like here to offer some hints.

Obviously, a crucial step in our construction was the introduction 
``by hand'' of up to 248 vector fields $B_\mu{}^\cM$ subject 
to the transformation rules
\ben
\Gd B_\mu{}^\cM ~=~ -2 \,\VV{\cM}{IJ}\,{\Beps}^I\psi^I_\mu
+\I \GG^I_{A\dA}\,\VV{\cM}{A}\,{\Beps}^I\Gg_\mu\chi^\dA \;.
\een
As mentioned before, for the 36 vector fields associated with the 36
commuting nilpotent directions in the $E_{8(8)}$ Lie algebra, this
formula can be derived directly from eleven dimensions
\cite{KoNiSa00}. Owing to the on-shell equivalence of vectors and
scalars, vector fields can be added with impunity in three dimensions,
but in extrapolating this step to eleven dimensions we seem to
run into an obstacle, because extra vector fields would normally
introduce new and unwanted propagating degrees of freedom. 
Nevertheless, the evidence for a generalized vielbein in eleven 
dimensions presented in \cite{deWNic86,Nico87,KoNiSa00}, and
the fact that a consistent gauging in three dimensions based on this
extrapolation does exist, prompt us to conjecture that {\it all} 248
vector fields introduced here have an eleven-dimensional origin. 
In \cite{KoNiSa00} it was observed that the physical bosonic degrees 
of freedom can be assembled into a 248-bein, which is just the lift 
of the $E_{8(8)}$ matrix $\cV$ to eleven dimensions. Assuming that 
there are indeed 248 vector fields, all bosonic fields would thus 
naturally fit into a (3+248)-bein
\ben
\left( 
\begin{array}{cc}
e_\mu{}^\Ga & B_\mu{}^{\cM}\,\cV_{\cM}{}^{\cA} \\[1ex]
0 & \cV_{\cM}{}^{\cA} 
\end{array}
\right) \;,
\een
which would also incorporate the three-form degrees of freedom and
would replace the original elfbein of $D\!=\!11$ supergravity
\ben
\left( 
\begin{array}{cc}
e_\mu{}^\Ga & B_\mu{}^m e_m{}^a \\[1ex]
0 & e_m{}^a 
\end{array}
\right) \;.
\een
The latter is just an element of the coset space $GL(11,{\mathbb
R})/SO(1,10)$ in a special gauge where the tangent space symmetry is
broken to $SO(1,2) \times SO(8)$. However, an analogous interpretation 
of the above (3+248)-bein remains to be found. Amongst other things, 
it would require replacing the action of the global $E_{8(8)}$ on
the 248-bein $\cV_\cM{}^\cA$ by some new type of general coordinate
transformations, in the same way as $GL(11)$ is 
replaced by diffeomorphisms in the vielbein description of 
Einstein's theory \cite{KoNiSa00}. The gauge groups found in the
compactification to three dimensions would then emerge as ``isometry
groups'' in a suitable sense. We also note that for the tangent space
group we have the embedding $SO(1,2) \times SO(16) \subset OSp(32)$, 
but there is no simple group generalizing $GL(11)$ that would contain 
$GL(3) \times E_{8(8)}$ and yield the right number of (bosonic) 
physical degrees of freedom upon division by $OSp(32)$ (see, 
however, \cite{West00} for an alternative ansatz based on the 
embedding $OSp(32) \subset OSp(64|1)$).

The challenge is therefore to find a reformulation of $D\!=\!11$
supergravity in terms of the above (3+248)-bein and an action, which
must still describe no more than 128 massless bosonic physical degrees
of freedom, despite the presence of new field components in eleven
dimensions. The only way to achieve this appears to be via a CS-like
action in eleven dimensions that would encompass all degrees of
freedom, and thus unify the Einstein-Hilbert and three-form actions of
the original theory\footnote{We are aware that the idea of
reformulating $D\!=\!11$ supergravity as a CS theory is not entirely
new. However, the present ansatz is evidently very different from
previous attempts in this direction.}.  In making these speculations
we are encouraged by the fact that, at least in three dimensions, the
dreibein $e_\mu{}^\alpha$, the gravitinos $\psi_\mu^I$ and the vector
fields are all governed by CS-type actions.

\newpage
\appendix
\renewcommand{\theequation}{\Alph{section}.\arabic{equation}}
\renewcommand{\thesection}{Appendix \Alph{section}:}

\mathon
\section{$\EE$ conventions}
\mathoff
\la{APE8}

The $E_{8(8)}$ generators $t^\cA$ are split into 120 compact ones
$X^{IJ}\equiv -X^{JI}$ and 128 noncompact ones $Y^A$, with $SO(16)$
vector indices $I, J, \dots \in\underline{16}$\,, spinor indices $A,
B, \dots\in\underline{128}$, and the collective labels $\CA, \CB,
\dots = ([IJ],A), \dots$. The
conjugate $SO(16)$ spinors are labeled by dotted indices $\dA, \dB,
\dots$.  In this $SO(16)$ basis the totally antisymmetric $E_{8(8)}$
structure constants $f^{\cA\cB\cC}$ possess the non-vanishing
components
\be
f^{I\!J,\,K\!L,\,M\!N} = 
-8\, \Gd\!\oversym{^{I[K}_{\vphantom{M}}\,\Gd_{MN}^{L]J}}
\;,\qquad
f^{I\!J,\,A,\,B}   = -\ft12 \GG^{IJ}_{AB} \;.\la{fABC}
\ee
$E_{8(8)}$ indices are raised and lowered by means of the
Cartan-Killing metric
\be
\eta^{\cA\cB}=\frac1{60} {\rm Tr} \, t^\cA t^\cB 
             =-\frac1{60} f^\cA{}_{\cC\cD}f^{\cB\cC\cD} \;,
\ee
with components $\eta^{AB}=\Gd^{AB}$ and
$\eta^{I\!J\,K\!L}=-2\Gd^{IJ}_{KL}$.  When summing over
antisymmetrized index pairs $[IJ]$, an extra factor of $\frac12$ is
always understood. Explicitly, the commutators are
\ba
\big[X^{IJ},X^{KL}\big] &=& 4\, \Gd\oversym{{}^{I[K}X^{L]J}\;}
\;,\non[.5ex]
\big[X^{IJ},Y^{A} \big] &=& -\ft12 \GG^{IJ}_{AB}Y^{B}
\;,\non[.5ex]
\big[Y^{A} ,Y^{B} \big] &=&  \ft14 \GG^{IJ}_{AB}X^{IJ}
\;.
\la{E8}
\ea

The equivalence of the fundamental and the adjoint representations 
of $E_{8(8)}$ plays an important role in our considerations; it is
expressed by the relation
\be\la{adjoint}
\cV^{-1} t^\cM\, \cV = {\cV^\cM}_\cA \,t^\cA \; 
\quad\Longleftrightarrow\quad
 \;
 {\cV^\cM}_\cA = \ft1{60} {\rm Tr} \, 
\big( t^\cM \, \cV \,t_\cA \,\cV^{-1} \big)\;.
\ee
Further formulas concerning the $E_{8(8)}$ Lie algebra, which will 
be used in this paper can be found in \ci{KoNiSa99,KoNiSa00}.

Let us finally point out that in the main text we use collective
labels $\CA, \CB, \dots$ and $\CM, \CN, \dots$ for the $\EE$ matrix
$\VV{\cM}{\cA}$ defined in \Ref{adjoint}, to distinguish the
transformation of these indices under the left and right action of
$\EE$ and $SO(16)$, respectively, according to \Ref{GHV}. Likewise,
$\GTh_{\cM\cN}$ is an $\EE$ tensor whereas $T_{\cA|\cB}$ transforms
under the local $SO(16)$, cf.~\Ref{defT}.

\newpage


\providecommand{\href}[2]{#2}
\begingroup\raggedright\endgroup

\end{document}